\documentclass[twocolumn,showpacs,preprintnumbers,amsmath,amssymb,
superscriptaddress,groupedaddress]{revtex4}

\usepackage{graphicx}
\usepackage{dcolumn}
\usepackage{bm}
\usepackage{amssymb}
\usepackage{color}

\begin{document}

\title{Numerical integration of variational equations}

\author{Ch.~Skokos} \affiliation{Max Planck Institute for the Physics
  of Complex Systems, N\"othnitzer Str.~38, D-01187 Dresden, Germany}
\author{E.~Gerlach} \affiliation{Lohrmann Observatory, Technical
  University Dresden, D-01062 Dresden, Germany}

\date{\today}

\begin{abstract}
  We present and compare different numerical schemes for the
  integration of the variational equations of autonomous Hamiltonian
  systems whose kinetic energy is quadratic in the generalized momenta
  and whose potential is a function of the generalized positions. We
  apply these techniques to Hamiltonian systems of various degrees of
  freedom, and investigate their efficiency in accurately reproducing
  well-known properties of chaos indicators like the Lyapunov
  Characteristic Exponents (LCEs) and the Generalized Alignment
  Indices (GALIs). We find that the best numerical performance is
  exhibited by the \textit{`tangent map (TM) method'}, a scheme based
  on symplectic integration techniques which proves to be optimal in
  speed and accuracy. According to this method, a symplectic
  integrator is used to approximate the solution of the Hamilton's
  equations of motion by the repeated action of a symplectic map $S$,
  while the corresponding tangent map $TS$, is used for the
  integration of the variational equations. A simple and systematic
  technique to construct $TS$ is also presented.
\end{abstract}

\maketitle

\section{Introduction}
\label{sec:intro}

Numerical integration is very often the only available tool for
investigating the properties of nonlinear dynamical systems. Different
numerical techniques \cite{Hairer_etal_93, Hairer_etal_02} have been
developed over the years which permit the fast and accurate time
evolution of orbits in such systems.

Of particular interest are the so-called \textit{`symplectic
  integrators'} which are numerical methods specifically aimed at
advancing in time the solution of Hamiltonian systems with the aid of
symplectic maps (see for example \cite[Chapt.~VI]{Hairer_etal_02},
\cite{SI_Ham} and references therein). Another challenging numerical
task in conservative Hamiltonian systems is to discriminate between
order and chaos. This distinction is a delicate issue because regular
and chaotic orbits are distributed throughout phase space in very
complicated ways. In order to address the problem several methods have
been developed, which can be divided into two major categories: the
ones based on the study of the evolution of deviation vectors from a
given orbit, like the computation of the maximal Lyapunov
Characteristic Exponent (mLCE) $\chi_1$ \cite{S10}, and those relying
on the analysis of the particular orbit itself, like the frequency map
analysis of Laskar \cite{FMA}.

Other chaos detection methods, belonging to the same category with the
evaluation of the mLCE, are the fast Lyapunov indicator (FLI)
\cite{FLI} and its variants \cite{Barrio}, the smaller alignment index
(SALI) \cite{S01b_SABV03_SABV04} and its generalization, the so-called
generalized alignment index (GALI) \cite{SBA07,SBA08}, and the mean
exponential growth of nearby orbits (MEGNO) \cite{MEGNO}. The
computation of these indicators require the numerical integration of
the so-called \textit{variational equations}, which govern the time evolution
of deviation vectors.

The scope of this paper is to present, analyze and compare
different numerical methods for the integration of the variational
equations. In our study we consider methods based  on symplectic
and non-symplectic integration techniques. The integration of the
variational equations by non-symplectic methods is straightforward
since one  simply has to integrate these equations simultaneously with
the equations of motion. This approach requires in general, more CPU
time than schemes based on symplectic
  integration  techniques for the same order of
accuracy and integration time step. For this reason we \emph{focus our
attention on methods based on symplectic schemes}, explaining in detail
their theoretical foundation and applying them to Hamiltonian systems
of different numbers of degrees of freedom.

The numerical solution of the variational equations obtained by the
various integration schemes studied are used for the computation of
the spectrum of the Lyapunov Characteristic Exponent (LCEs) and the
GALIs. We chose to compute these two chaos indicators among the
indices based on the evolution of deviation vectors, because the
computation of the mLCE is the elder and most commonly employed chaos
detection technique, while the computation of the whole spectrum of
LCEs and GALIs requires the evolution of more than one deviation
vector and thus is strongly influenced by inaccuracies of the
integration procedure. We investigate the numerical efficiency of the
different integration methods by comparing the CPU times they require
for the computation of the LCEs and the GALIs, as well as their
accuracy in reproducing well-known properties of these chaos
indicators. In particular, we check whether the set of computed LCEs
consists of pairs of values having opposite signs, and if the time
evolution of GALIs follows specific theoretically predicted laws.

The paper is organized as follows: after introducing the concept of
variational equations in the next section, we describe in sections
\ref{sec:LCEs} and \ref{sec:GALI} the LCEs and the GALIs respectively,
which are the two chaos indicators we use in our study. Then, in section
\ref{sec:SI} we give the basic properties of symplectic
integrators. Section \ref{sec:num_int} is devoted to the detailed
description of several numerical schemes for the integration of the
variational equations of Hamiltonian systems. Applications of these
schemes to regular and chaotic orbits of systems with two or more
degrees of freedom are presented in section \ref{sec:numerics}, where
also the efficiency of each technique is discussed. Finally, in
section \ref{sec:summary}, we summarize the results and present our
conclusions, while in the appendix the explicit expressions of the
various integration methods for the H\'enon-Heiles system are
given.

\section{The variational equations}

Let us consider an \textit{autonomous Hamiltonian system} of $N$
degrees of freedom ($N$D) having a Hamiltonian function
\begin{equation}
  H(q_1,q_2, \ldots, q_N,p_1,p_2, \ldots, p_N)=h=\mbox{constant},
\label{eq:Ham_gen}
\end{equation}
where $q_i$ and $p_i$, $i=1,2,\ldots,N$ are the generalized
coordinates and conjugate momenta respectively. An orbit in the
$2N$-dimensional phase space $\mathcal{S}$ of this system is defined
by the vector
\begin{equation}
  \vec{x}(t)=(q_1(t),q_2(t), \ldots, q_N(t),p_1(t),p_2(t), \ldots, p_N(t)),
\label{eq:x_vect}
\end{equation}
with $x_i=q_i$, $x_{i+N}=p_i$, $i=1,2,\ldots,N$. The time evolution of
this orbit is governed by the \textit{Hamilton's equations of motion},
which in matrix form are given by
\begin{equation}
  \dot{\vec{x}}= \left[ \begin{array}{cc} \frac{\partial
        H}{\partial \vec{p}}\,\, & - \frac{\partial H}{\partial \vec{q}}
\end{array} \right]^{\mathrm{T}}= \textbf{J}_{2N}\cdot\textbf{D}_H,
\label{eq:Hameq_gen}
\end{equation}
with $ \vec{q}=(q_1(t),q_2(t), \ldots, q_N(t))$, $
\vec{p}=(p_1(t),p_2(t), \ldots, p_N(t))$, and
\[
\textbf{D}_H=\left[ \begin{array}{cccccccc} \frac{\partial H}{\partial
      q_1} &\frac{\partial H}{\partial q_2}& \cdots & \frac{\partial
      H}{\partial q_N} & \frac{\partial H}{\partial p_1}
    &\frac{\partial H}{\partial p_2}& \cdots & \frac{\partial
      H}{\partial p_N}
\end{array}
\right]^{\mathrm{T}}
\]
with $(^{\mathrm{T}})$ denoting the transpose matrix. Matrix
$\textbf{J}_{2N}$ has the following block form
\[
\textbf{J}_{2N}= \left[ \begin{array}{cc} \textbf{0}_{N} & \textbf{I}_{N} \\
    -\textbf{I}_{N} & \textbf{0}_{N}
\end{array}
 \right] ,
\]
with $\textbf{I}_{N}$ being the $N\times N$ identity matrix and
$\textbf{0}_{N}$ being the $N\times N$ matrix with all its elements
equal to zero.

An initial deviation vector $\vec{w}(0)=(\delta x_1(0),\delta
x_2(0),\ldots, \delta x_{2N}(0)) $ from an orbit $\vec{x}(t)$ evolves
in the \textit{tangent space} $\mathcal{T}_{\vec{x}} \mathcal{S}$ of
$\mathcal{S}$ according to the so-called \textit{variational
  equations}
\begin{equation}
  \dot{\vec{w}} =\left[
    \textbf{J}_{2N}\cdot\textbf{D$^2_H$}(\vec{x}(t)) \right] \cdot\vec{w}=:
  \textbf{A}(t) \cdot\vec{w} \, ,
\label{eq:var}
\end{equation}
with $\textbf{D$^2_H$}(\vec{x}(t))$ being the Hessian matrix of
Hamiltonian (\ref{eq:Ham_gen}) calculated on the reference orbit
$\vec{x}(t)$, i.~e.
\[
\textbf{D$^2_H$}(\vec{x}(t))_{i,j} = \left. \frac{\partial^2 H}{\partial x_i
    \partial x_j}\right|_{\vec{x}(t)} \,\,\,,\,\,\, i,j=1,2,\ldots,2N.
\]
Equations (\ref{eq:var}) are a set of linear differential equations
with respect to $\vec{w}$, having time dependent coefficients since
matrix $\textbf{A}(t)$ depends on the particular reference orbit,
which is a function of time $t$.

In the present paper we consider   autonomous Hamiltonians
of the form
\begin{equation}
  H(\vec{q},\vec{p} )= \frac{1}{2} \sum_{i=1}^{N} p_i^2+ V(\vec{q}),
\label{eq:HH_usual}
\end{equation}
with $V(\vec{q})$ being the potential function. The Hamilton's equations
of motion (\ref{eq:Hameq_gen}) become
\begin{equation}
\dot{\vec{x}}=\left[ \begin{array}{c}
\dot{\vec{q}} \\
\dot{\vec{p}}
\end{array} \right] =
\left[ \begin{array}{c}
 \ \vec{p} \\
-\frac{\partial V(\vec{q})}{\partial \vec{q}}
\end{array} \right],
\label{eq:eq_mot_V}
\end{equation}
while the variational equations (\ref{eq:var}) of this system take the
form
\begin{equation}
\begin{array} {c}
  \displaystyle
  \dot{\vec{w}}=\left[ \begin{array}{c}
      \dot{\vec{\delta q}} \\
      \dot{\vec{\delta p}}
\end{array} \right] = \textbf{A}(t) \cdot\vec{w}= \left[
\begin{array}{cc}
  \textbf{0}_{N} & \textbf{I}_{N} \\
  -\textbf{D$^2_V$}(\vec{q}(t)) & \textbf{0}_{N}
\end{array}\right] \cdot \left[ \begin{array}{c}
  \vec{\delta q} \\
  \vec{\delta p}
\end{array} \right]\Rightarrow \\ \\
\begin{array}{l}\displaystyle
  \dot{\vec{\delta q}}=\vec{\delta p} \\ \displaystyle
  \dot{\vec{\delta p}} =- \mathbf{D}^2_V(\vec{q}(t))\vec{\delta q}
\end{array}
\end{array}
\label{eq:w_V}
\end{equation}
with $ \vec{\delta q}=(\delta q_1(t),\delta q_2(t), \ldots, \delta
q_N(t))$, $ \vec{\delta p}=(\delta p_1(t),\delta p_2(t) \ldots, \delta
p_N(t))$, and
\begin{equation}
  \textbf{D$^2_V$}(\vec{q}(t))_{jk}=\left. \frac{\partial^2
      V(\vec{q})}{\partial q_j \partial q_k} \right|_{\vec{q}(t)}\,\,\, ,\,\,\,
  j,k=1,2,\ldots,N.
\label{eq:D2V}
\end{equation}
Thus, the tangent dynamics of Hamiltonian (\ref{eq:HH_usual}) is
represented by the time dependent Hamiltonian function
\begin{equation}
  H_V( \vec{ \delta q}, \vec{\delta p};t)=\frac{1}{2} \sum_{i=1}^{N} \delta
  p_i^2+ \frac{1}{2}\sum_{j,k}^{N} \textbf{D$^2_V$}(\vec{q}(t))_{jk} \delta q_j
  \delta q_k ,
\label{eq:HV_general}
\end{equation}
which we call the \textit{`tangent dynamics Hamiltonian' (TDH)}, and
whose equations of motion are exactly the variational equations
(\ref{eq:w_V}).

\section{The Lyapunov Characteristic Exponents}
\label{sec:LCEs}

The LCEs are asymptotic measures characterizing the average rate of
growth (or shrinking) of small perturbations to the solutions of a
dynamical system. Their concept was introduced by Lyapunov when
studying the stability of non-stationary solutions of ordinary
differential equations \cite{L1892}, and has been widely employed in
studying dynamical systems since then. A detailed review of the theory
of the LCEs, as well as of the numerical techniques developed for
their computation can be found in \cite{S10}.

The theory of LCEs was applied to characterize chaotic orbits by
Oseledec \cite{O68}, while the connection between LCEs and exponential
divergence of nearby orbits was given in \cite{BGS76,P77}. For a
chaotic orbit at least one LCE is positive, implying exponential
divergence of nearby orbits, while in the case of regular orbits all
LCEs are zero or negative. Therefore, the computation of the mLCE
$\chi_1$ is sufficient for determining the nature of an orbit, because
$\chi_1>0$ guarantees that the orbit is chaotic.

The mLCE is computed as the limit for $t\rightarrow \infty$ of the
quantity
\begin{equation}
X_1(t) =\frac{1}{t} \ln \frac{\| \vec{w}(t) \|}{\|
\vec{w}(0) \|},
\label{eq:X_1_t}
\end{equation} 
often called \textit{finite time mLCE}, where $\vec{w}(0)$,
$\vec{w}(t)$ are deviation vectors from a given orbit, at times $t=0$
and $t>0$ respectively, and $\| \,\cdot\, \|$ denotes the norm of a
vector. So, we have
\begin{equation}
\chi_1= \lim_{t\rightarrow \infty} X_1(t).
\label{eq:chi_1_t_limit}
\end{equation} 
If the energy surface defined by (\ref{eq:Ham_gen}) is compact, it has
been shown that this limit is finite, independent of the choice of the
metric for the phase space and converges to $\chi_1$ for almost all
initial vectors $\vec{w}(0)$ \cite{O68,BGGS80a,BGGS80b}. $X_1(t)$
tends to zero in the case of regular orbits following a power law
\cite{BGS76}
\begin{equation}
X_1(t) \propto t^{-1},
\label{eq:X1_t}
\end{equation}
while it tends to nonzero values in the case of chaotic orbits.

An $N$D Hamiltonian system has $2N$ (possibly non-distinct) LCEs,
which are ordered  as $\chi_1 \geq \chi_2 \geq \cdots \geq
\chi_{2N}$. In \cite{BGGS78} a theorem was formulated, which led
directly to the development of a numerical technique for the
computation of all LCEs, based on the time evolution of many
deviation vectors,  kept linearly independent through a
Gram-Schmidt orthonormalization procedure. The theoretical framework,
as well as the corresponding numerical method for the computation of
all LCEs (usually called the \textit{`standard method'}), were
given in \cite{BGGS80a,BGGS80b}. According to this method all other
LCEs $\chi_2$, $\chi_3$ etc., apart from the mLCE obtained from
(\ref{eq:chi_1_t_limit}), are computed as the limits for $t
\rightarrow \infty$ of some appropriate quantities $X_2(t)$, $X_3(t)$
etc., which are called the \textit{finite time LCEs} (see
\cite{BGGS80b,S10} for more details).  We note that throughout the
present paper, whenever we need to compute the values of the LCEs, we
apply the discrete QR-decomposition technique
\cite[Sect.~2.10]{NumRec}, which is a variation of the standard method
(see Sect.~6.3 of \cite{S10} for more details).

It has been shown in \cite{BGGS80a} that in the case of an autonomous
Hamiltonian flow, the set of LCEs consists of pairs of values having
opposite signs
\begin{equation}
\chi_i= - \chi_{2N-i+1} \,\,\, , \,\,\, i=1,2,\ldots, N.
\label{eq:op_signs}
\end{equation} 
In addition, since the Hamiltonian function is an integral of motion,
at least two LCEs vanish, i.~e.
\begin{equation}
\chi_N= \chi_{N+1}=0,
\label{eq:spct_00}
\end{equation} 
while the presence of any additional independent integral of motion
leads to the vanishing of another pair of LCEs.

\section{The Generalized Alignment Index}
\label{sec:GALI}

The GALI is an efficient chaos detection technique introduced in
\cite{SBA07} as a generalization of a similar indicator called the
smaller alignment index (SALI) \cite{S01b_SABV03_SABV04}. The method
has been applied successfully for the discrimination between regular
and chaotic motion, as well as for the detection of regular motion on
low dimensional tori to different dynamical systems \cite{SBA08,
  GALI_appl}.

The GALI of order $k$ ($G_k$) is determined through the evolution
of $2 \leq k \leq 2N$ initially linearly independent deviation vectors
$\vec{w}_i(0)$, $i=1,2,\ldots,k$. The time evolution of each deviation
vector is governed by the variational equations (\ref{eq:w_V}). Each
evolved deviation vector $\vec{w}_i(t)$ is normalized from time to
time, having its norm equal to 1, in order to avoid overflow problems,
but its direction is left intact. Then, according to \cite{SBA07},
$G_k$ is defined to be the volume of the $k$-parallelogram having
as edges the $k$ unitary deviation vectors $\hat{w}_i(t)$,
$i=1,2,\ldots,k$. This volume is equal to the norm of the wedge
product of these vectors, and $G_k$ is given by
\begin{equation}
  G_k(t)=\| \hat{w}_1(t)\wedge \hat{w}_2(t)\wedge \cdots
  \wedge\hat{w}_k(t) \|.
\label{eq:GALI}
\end{equation}
From this definition it is evident that if at least two of the
deviation vectors become linearly dependent, the wedge product in
(\ref{eq:GALI}) becomes zero and the $G_k$ vanishes.

Expanding the wedge product (\ref{eq:GALI}) into a sum of determinants
and studying the asymptotic behavior of those who vary the
\textit{slowest} in time, it is possible to show analytically the
following \cite{SBA07}: in the case of a chaotic orbit all deviation
vectors tend to become linearly dependent, aligning in the direction
defined by the mLCE and $G_k$ tends to zero exponentially
following the law
\begin{equation}
  G_k(t) \propto e^{-\left[ (\sigma_1-\sigma_2) + (\sigma_1-\sigma_3)+
      \cdots+ (\sigma_1-\sigma_k)\right]t},
\label{eq:GALI_chaos}
\end{equation}
where $\sigma_1, \ldots, \sigma_k$ are approximations of the first $k$
largest Lyapunov exponents.
On the other hand, in the case of regular motion on an $N$-dimensional
torus, all deviation vectors tend to fall on the $N$-dimensional
tangent space of this torus. Thus, if we start with $k\leq N$ general
deviation vectors they will remain linearly independent on the
$N$-dimensional tangent space of the torus, since there is no
particular reason for them to become aligned. As a consequence
$G_k$ is different from zero and remains practically constant for
$k\leq N$. On the other hand, $G_k$ tends to zero for $k>N$,
since some deviation vectors will eventually become linearly
dependent, following a particular power law which depends on the
dimensionality $N$ of the torus and the number $k$ of deviation
vectors. The behavior of $G_k$ for regular orbits lying on
$N$-dimensional tori is given by
\begin{equation}
  G_k (t) \propto \left\{ \begin{array}{ll} \mbox{constant} & \mbox{if
        $2\leq k \leq N$} \\ \frac{1}{t^{2(k-N)}} & \mbox{if $N< k \leq 2N$} \\
\end{array}\right. .
\label{eq:GALI_order_all_N}
\end{equation}
If the regular orbit lies on a low dimensional torus, i.~e. an
$s$-dimensional torus with $2\leq s \leq N$ then $G_k$ remains
practically constant and different from zero for $k\leq s$ and tends
to zero for $k>s$ following particular power laws (see \cite{SBA08}
for more details).

In order to compute the value of $G_k$ we consider the $2N \times
k$ matrix $\textbf{W}(t)$ having as columns the coordinates
$w_{ji}(t)$ of the unitary deviation vectors $\hat{w}_i(t)$,
$i=1,2,\ldots,k$, $j=1,2,\ldots,2N$, with respect to the usual
orthonormal basis $\hat{e}_1= (1,0,0,\ldots,0)$, $\hat{e}_2=
(0,1,0,\ldots,0)$, ..., $\hat{e}_{2N}= (0,0,0,\ldots,1)$ of the
$2N$-dimensional tangent space $\mathcal{T}_{\vec{x}} \mathcal{S}$ and
perform the Singular Value Decomposition (SVD) of this matrix. Then,
as it was shown in \cite{SBA08}, $G_k$ is equal to the product of
the singular values $z_i$, $i=1,2,\ldots,k$ of matrix $\textbf{W}(t)$, i.~e.
\begin{equation}
G_k(t) = \prod_{i=1}^k z_i(t)\, .
\label{eq:gali_svd}
\end{equation}

\section{Symplectic integrators}
\label{sec:SI}


Let us discuss in some detail how we can integrate the equations of
motion (\ref{eq:Hameq_gen}) of a general Hamiltonian
(\ref{eq:Ham_gen}) by a symplectic integration scheme, focusing our
attention on a particular family of integrators presented in
\cite{LR01}. Defining the Poisson bracket  of functions
$f(\vec{q},\vec{p})$, $g(\vec{q},\vec{p})$ by \cite{comment1}:
\begin{equation}
\{ f,g\}=\sum_{l=1}^{N} \left( \frac{\partial f}{\partial q_l}
\frac{\partial g}{\partial p_l}  - \frac{\partial f}{\partial p_l}
\frac{\partial g}{\partial q_l}\right) , \label{eq:Poisson}
\end{equation}
the Hamilton's equations of motion (\ref{eq:Hameq_gen}) take the form
\begin{equation}
\frac{d \vec{x}} {dt}= \{\vec{x},H\} = L_H \vec{x},
\label{eq:Hameq_x}
\end{equation}
where $L_H$ is the differential operator defined by $L_{\chi}f=\{f,
 \chi\}$.  The solution of Eq.~(\ref{eq:Hameq_x}), for initial conditions
$\vec{x}(0)=\vec{x}_0$, is formally written as
\begin{equation}
\vec{x}(t)=\sum_{n\geq 0} \frac{t^n}{n!} L_H^n \vec{x}_0=e^{t L_H}
\vec{x}_0.
\label{eq:Ham_sol}
\end{equation}

Let us assume that the Hamiltonian function $H$ can be split into two
integrable parts as $H=A+B$. A symplectic scheme for integrating
equations (\ref{eq:Hameq_x}) from time $t$ to time $t+\tau$ consists
of approximating, in a symplectic way, the operator $ e^{\tau L_H}=
e^{\tau(L_A+L_{B} ) } $ by an integrator of $j$ steps involving
products of operators $e^{c_i \tau L_A}$ and $e^{d_i \tau L_{B}}$,
$i=1,2,\ldots, j$, which are exact integrations over times $c_i \tau$
and $d_i \tau$ of the integrable Hamiltonians $A$ and $B$. The
constants $c_i$, $d_i$, which in general can be positive or negative,
are chosen to increase the order of the remainder of this
approximation. So $e^{\tau L_A}$, $e^{\tau L_{B}}$ are actually
symplectic maps acting on the coordinate vector $\vec{x}$. Therefore
the integration of equations (\ref{eq:Hameq_x}) over one time step
$\tau$, which evolves the initial coordinate vector $\vec{x}(t)$ to
its final state $\vec{x}(t+\tau)$, is represented by the action on
$\vec{x}(t)$ of a symplectic map $S$ produced by the composition of
products of $e^{c_i \tau L_A}$ and $e^{d_i \tau L_{B}}$. In this
context several symplectic integrators of different orders have been
developed by various researchers \cite{SI_methods,ML95}.

In \cite{LR01} the families of SBAB (and SABA) symplectic integrators,
which involve only forward (positive) integration steps were
introduced. These integrators were adapted for the integration of
perturbed Hamiltonians of the form $H=A+\epsilon B$, where both $A$
and $B$ are integrable and $\epsilon$ is a small parameter. A
particular integrator SBAB$_n$ ($S_{Bn}$), or SABA$_n$ ($S_{An}$), involves $n$ steps, i.~e.
$n$ applications of products of $e^{c_i \tau L_A}$ and $e^{d_i \tau
  L_{\epsilon B}}$, and is of order $\mathcal{O}(\tau^{2n}\epsilon
+\tau^{2}\epsilon^2)$ with respect to the integration step
$\tau$. This means that by using these integrators, we are actually
approximating the dynamical behavior of the real Hamiltonian
$A+\epsilon B$ by a Hamiltonian $H^* = A+\epsilon B +
\mathrm{\cal{O}}(\tau^{2n} \epsilon +\tau^2 \epsilon ^2)$, i.~e.~we
introduce an error term of the order $\tau^{2n} \epsilon +\tau^2
\epsilon ^2$.

The accuracy of the $S_{Bn}$ ($S_{An}$) integrator can be improved
when the commutator term $C=\{B,\{B,A\}\}$ \cite{comment2} leads to an
integrable system, as in the common situation of $A$ being quadratic
in momenta $\vec{p}$ and $B$ depending only on positions $\vec{q}$. In
this case, two corrector terms of small backward (negative) steps can
be added to the integrator $S_{Bn}$
\begin{equation}
S_{Bn}^c= e^{- \tau^3 \epsilon^2 \frac{g}{2}
L_{C}}(S_{Bn})e^{- \tau^3 \epsilon^2 \frac{g}{2} L_{C}}.
\label{eq:saba2c}
\end{equation}
A similar expression is valid also for $S_{An}$. The value of constant
$g$ is chosen in order to eliminate the $\tau^2 \epsilon^2$ dependence
of the remainder which becomes of order $\mathcal{O}(\tau^{2n}
\epsilon +\tau^4 \epsilon ^2)$. The SBAB (SABA) integrators have
already proved to be very efficient for the numerical study of
different dynamical systems \cite{LR01,si_appl1,si_appl2}. We note
that several authors have used commutators for improving the
efficiency of symplectic integrators (e.~g.~\cite{Chin97,OMF_23}).

Setting $\epsilon=1$ we can apply the SBAB (SABA) integration schemes
for the integration of Hamiltonian (\ref{eq:HH_usual}), since this
Hamiltonian can be written as $H=A+B$, with
\begin{equation}
A(\vec{p})  =  \displaystyle \frac{1}{2} \sum_{i=1}^{N} p_i^2, \,\,\,\,\, B(
\vec{q})  = 
\displaystyle V(\vec{q}),
\label{eq:H_AB}
\end{equation}
being both integrable. The maps $e^{\tau L_{A}}$, $e^{\tau L_{B}}$,
which propagate the set of initial conditions $(\vec{q},\vec{p})$ at
time $t$, to their final values $(\vec{q}\,',\vec{p}\,')$ at time
$t+\tau$, for the Hamiltonian functions $A(\vec{p})$ and $B(\vec{q})$
(\ref{eq:H_AB}) are
\begin{equation}
e^{\tau L_{A}}: \left\{ \begin{array}{lll}
\vec{q}\,' & = & \vec{ q} + \vec{p} \tau \\
\vec{p}\,' & = & \vec{p}
\end{array}\right. ,
\label{eq:H_LA}
\end{equation}
and
\begin{equation}
e^{\tau L_{B}}: \left\{
\begin{array}{lll}
\vec{q}\,'  & = &  \vec{q}\\
\vec{p}\,' & = &  \displaystyle \vec{p} - \frac{\partial V( \vec{ q})}{\partial
\vec{ q}}   \tau
\end{array}
\right. 
\label{eq:H_LB}
\end{equation}
respectively.  For Hamiltonian (\ref{eq:HH_usual}) the corrector term
is given by
\begin{equation}
C=\{B,\{B,A\}\}= \sum_{i=1}^N \left(\frac{\partial V\left(\vec{
q}\right)}{\partial \vec{q_i}} \right)^2 ,
\label{eq:H_ABB}
\end{equation}
which is a function of only the coordinates $\vec{q}$ and thus easily
integrated as
\begin{equation}
e^{\tau L_{C}}: \left\{ \begin{array}{lll} \vec{ q}' & = & \vec{ q}\\ \vec{p}\,'
& = & \displaystyle \vec{ p} - \frac{\partial C(
\vec{ q})} {\partial \vec{ q}}  \tau
\end{array}\right. .
\label{eq:H_LC}
\end{equation}
In Appendix \ref{ap:HH} we give the explicit formulas of equations
(\ref{eq:H_LA}), (\ref{eq:H_LB}) and (\ref{eq:H_LC}) for the
H\'enon-Heiles system (\ref{eq:HH}).

\section{Numerical integration of variational equations}
\label{sec:num_int}

In this section we present several numerical schemes for the
integration of the variational equations, considering both
non-symplectic techniques and methods based on symplectic
integrators. The latter schemes are quite general and any symplectic
integrator can be used for their implementation. In our study we
consider an efficient fourth order symplectic integrator, the
$S_{B2}^c$ \cite{Chin97,LR01}, which has an extra degree of complexity
with respect to integrators composed of products of maps $e^{\tau
  L_{A}}$, and $e^{\tau L_{B}}$, since it requires the application of
the corrector term $C$ (\ref{eq:H_ABB}).

\subsection{Non-symplectic schemes}

In order to follow the evolution of a deviation vector, the
variational equations (\ref{eq:w_V}) have to be integrated
simultaneously with the Hamilton's equations of motion
(\ref{eq:eq_mot_V}), since matrix $ \mathbf{D}^2_V(t)$ depends
on the particular reference orbit $\vec{x}(t)$, which is a solution of
equations (\ref{eq:eq_mot_V}). Any non-symplectic numerical
integration algorithm can be used for the integration of the whole set
of equations (\ref{eq:eq_mot_V}) and (\ref{eq:w_V}).

In our study we use the DOP853 integration method which has been
proven to be very efficient. The DOP853 integrator \cite{foot1} is an
explicit non-symplectic Runge-Kutta integration scheme of order 8,
based on the method of Dormand and Price (see
\cite[Sect.~II.5]{Hairer_etal_93}).  Two free parameters, $\tau$ and
$\delta$, are used to control the numerical performance of the
method. The first one defines the time span between two successive
outputs of the computed solution.  After each step of length $\tau$
the values of LCEs (GALIs) are computed and the deviation vectors are
orthonormalized (normalized).  For the duration of each step $\tau$,
the integrator adjusts its own internal time step, so that the local
one-step error is kept smaller than the user-defined threshold value
$\delta$. For DOP853 the estimation of this local error and the step
size control is based on embedded formulas of orders 5 and 3.

\subsection{Integration of the tangent dynamics Hamiltonian}
\label{sec:TDH}

Another approach to compute the evolution of deviation vectors is to
initially integrate the Hamilton's equations of motion
(\ref{eq:eq_mot_V}), in order to obtain the time evolution of the
reference orbit $\vec{x}(t)$, and then to use this numerically known
solution for solving the equations of motion of the TDH 
(\ref{eq:HV_general}), which are actually the variational equations
(\ref{eq:w_V}).

In practice one numerically solves the Hamilton's equations of motion
(\ref{eq:eq_mot_V}) by any (symplectic or non-symplectic) integration
scheme to obtain the values $\vec{x}(t_i)$ at $t_i=i\, \Delta t$,
$i=0,1,2,\ldots$, where $\Delta t$ is the integration time step of
these orbits.  Of course, the accuracy of the particular numerical
scheme used for the construction of the time series $\vec{x}(t_i)$
will affect the quality of the numerical solution of the variational
equations, regardless of the numerical scheme used for solving them.
Having computed the values $\vec{x}(t_i)$ different methods can be
applied for approximating the solution of the variational equations,
which will be discussed in the following sections.

\subsubsection{TDH with piecewise constant coefficients}
\label{sec:TDH_const}

One method is to approximate the actual time dependent TDH
(\ref{eq:HV_general}) by a Hamiltonian with piecewise constant
coefficients. This means to assume that the coefficients
$\textbf{D$^2_V$}(\vec{q}(t))_{jk}$ $j,k=1,2,\ldots, N$ of $H_V$
(\ref{eq:HV_general}) are constants equal to
$\textbf{D$^2_V$}(\vec{q}(t_i))_{jk}$ for the time interval $\left[
  t_i, t_i+\Delta t\right)$. These constants are determined by the
values of the orbit's coordinates and are known, since we know the
time series $\vec{x}(t_i)=\left( \vec{q}(t_i), \vec{p}(t_i)
\right)$. Thus, for each time interval $[t_i, t_i+\Delta t)$ we end up
with a quadratic form Hamiltonian function $H_V( \vec{ \delta q},
\vec{\delta p};t_i)$, whose equations of motion form a linear system
of differential equations with constant coefficients.

The Hamiltonian $H_V( \vec{ \delta q}, \vec{\delta p};t_i)$ can be
integrated by any symplectic or non-symplectic integration scheme, or
can be explicitly solved by performing a canonical transformation to
new variables $\vec{Q}$, $\vec{P}$, so that the transformed
Hamiltonian $H_{VQP}$ becomes a sum of uncoupled 1D Hamiltonians,
whose equations of motion can be integrated immediately.  To this end,
let $\lambda_k$ be the eigenvalues and $\vec{v}_k$, $k=1,2,\ldots,N$
the unitary eigenvectors of the constant matrix
$\textbf{D$^2_V$}(\vec{q}(t_i))$. Then matrix $\mathbf{T}$, having as
columns the eigenvectors $\vec{v}_k$, defines a canonical change of
variables $\vec{q}=\mathbf{T} \vec{Q}$, $\vec{p}=\mathbf{T} \vec{P}$,
which gives $H_V$ the diagonal form
\begin{equation}
H_{VQP}=\sum_{i=1}^N \frac{1}{2} \left( P_i^2+ \lambda_i Q_i^2\right).
\label{eq:HVQP}
\end{equation}
The equations of motion of $H_{VQP}$ are then easily solved.

In our study we use the same symplectic integrator ($S_{B2}^c$) both
for obtaining the time series $\vec{x}(t_i)$ and for integrating the
quadratic form Hamiltonian $H_V( \vec{ \delta q}, \vec{\delta p};t_i)$
in the time interval $\left[ t_i, t_i+\Delta t\right)$. We name this
approach the \textit{TDHcc method} (cc: constant coefficients). An
alternative approach is to compute the exact solution of the equations
of motion of $H_V( \vec{ \delta q}, \vec{\delta p};t_i)$ (whose
piecewise constant coefficients are obtained by the symplectic
integration of the orbit using the $S_{B2}^c$ scheme) by transforming
it to a system of $N$ uncoupled harmonic oscillators through the
canonical transformation induced by matrix $\mathbf{T}$. This approach
is called the \textit{TDHes method} (es: exact solution).

In general, the transformation matrix $\mathbf{T}$ is determined for
each time interval $\left[t_i, t_i+\Delta t\right)$ by solving
numerically the eigenvalue problem
\begin{equation}
\mathbf{D}^2_V(\vec{q}(t_i))\,\vec{v}=\lambda \vec{v},
\label{eq:eigen_pr}
\end{equation}
a procedure which could become computationally very time consuming,
especially for systems with many degrees of freedom. On the other
hand, in some simple low dimensional cases, like for example the
H\'enon-Heiles system (\ref{eq:HH}), the transformation matrix
$\mathbf{T}$ can be determined analytically (see Appendix
\ref{ap:transform}).

\subsubsection{Integration of the TDH in an extended phase space}
\label{sec:EPS}

Instead of approximating  $H_V$ (\ref{eq:HV_general}) by a
quadratic form having constant coefficients for each time interval
$\left[ t_i, t_i+\Delta t\right)$, we can explicitly treat $H_V$ as a
time dependent Hamiltonian. This time dependency is due to the fact
that the coefficients of $H_V$ are functions of the orbit's
coordinates $\vec{q}(t)$. Like in the previous approach, we consider the
time series $\vec{q}(t_i)$ to be known from the numerical integration
of the Hamilton's equations (\ref{eq:eq_mot_V}).

The $N$D time dependent Hamiltonian $H_V$ can be transformed to a time
independent Hamiltonian $\widetilde{H}_V$ with an extra degree of
freedom by considering the time $t$ as an additional coordinate (see
for example \cite[Sect.~1.2b]{LichtenbergL_1992}). For this purpose,
we add to the Hamilton's equations of motion of $H_V$ the equations
\begin{equation}
\dot{t}=1\,\,\, , \,\,\, \dot{H_V}=\frac{\partial H_V}{\partial t}.
\label{eq:tH}
\end{equation}
Then we set $t$ and $-H_V$ as an additional coordinate and momentum
respectively, i.~e.~$\delta q_{N+1}=t$, $\delta p_{N+1}=-H_V$, and
define the new Hamiltonian
\begin{equation}
\widetilde{H}_V (\vec{\xi},\vec{\eta})=H_V(\vec{\delta q}, \vec{\delta
p};t)+\delta p_{N+1},
\label{eq:H_N+1}
\end{equation}
where $\vec{\xi}=(\vec{\delta q},t)$ and $\vec{\eta}=(\vec{\delta
  p},-H_V)$ are respectively the new coordinates and momenta. The flow
in the $(2N+2)$-dimensional extended phase space of the $(N+1)$D
Hamiltonian $\widetilde{H}_V$ is parameterized by a `new' time $\zeta$
such that $t(\zeta)=\zeta$, which does not appear explicitly in the
functional form of $\widetilde{H}_V$ (\ref{eq:H_N+1}). The set of
equations (\ref{eq:w_V}) and (\ref{eq:tH}) are the Hamilton's
equations of motion of $\widetilde{H}_V$.

The dynamics of the $N$D TDH $H_V$ (\ref{eq:HV_general}) is equivalent
to that of the $(N+1)$D Hamiltonian
\begin{equation}
\begin{array}{c}
\displaystyle \widetilde{H}_V( \vec{ \delta q},t, \vec{\delta p}, p_{N+1}) =
\frac{1}{2} \sum_{j=1}^{N} \delta p_i^2 + \delta p_{N+1} + \\ \\ \displaystyle
+ \frac{1}{2}\sum_{j,k}^{N} \textbf{D$^2_V$}(\vec{q}(t))_{jk} \delta q_j \delta
q_k.
\end{array}
\label{eq:HV_tilde_general}
\end{equation}
This Hamiltonian can be easily integrated by any symplectic
integration scheme, since it can be split into two integrable parts
\begin{equation}
\begin{array}{lll}
\widetilde{A}(\vec{\delta p}, \delta p_{N+1})& = & \displaystyle \frac{1}{2}
\sum_{j=1}^{N} \delta p_i^2 + \delta p_{N+1},\\ \\ \widetilde{B}( \vec{ \delta
q},t) & = & \displaystyle \frac{1}{2}\sum_{j,k}^{N}
\textbf{D$^2_V$}(\vec{q}(t))_{jk} \delta q_j \delta q_k.
\end{array}
\label{eq:HV_tilde_AB}
\end{equation}
The maps $e^{\tau L_{\widetilde{A}}}$, $e^{\tau L_{\widetilde{B}}}$,
which propagate the set of initial conditions $(\vec{\delta
  q},t,\vec{\delta p},\delta p_{N+1})$ at time $t$, to their final
values $(\vec{\delta q}',t',\vec{\delta p}',\delta p'_{N+1})$ at time
$t+\tau$ are
\begin{equation}
e^{\tau L_{\widetilde{A}}}: \left\{ \begin{array}{lll}
\vec{\delta q}' & = &  \vec{\delta q} + \vec{\delta p} \tau\\
t' & = & t + \tau \\
\vec{\delta p}' & = & \vec{\delta p} \\
\delta p'_{N+1} & = & \delta p_{N+1}
\end{array}\right. ,
\label{eq:HV_tilde_LA}
\end{equation}
\begin{equation}
e^{\tau L_{\widetilde{B}}}: \left\{ \begin{array}{lll} \vec{\delta q}' & = &
\vec{\delta q}\\ t' & = & t\\ \vec{\delta p}' & = & \displaystyle \vec{\delta
p} - \frac{\partial \widetilde{B}(\vec{
\delta q},t)}{\partial \vec{\delta q}}  \tau \\ \delta p'_{N+1} & = & \displaystyle \delta p_{N+1} -
\frac{\partial \widetilde{B}(\vec{ \delta q},t)}{\partial t}   \tau
\end{array}\right. .
\label{eq:HV_tilde_LB}
\end{equation}
The corrector term of the SBAB and SABA integration schemes
\begin{equation}
\widetilde{C}=\{\widetilde{B},\{\widetilde {B}, \widetilde{A}\}\}=\sum_{i=1}^N
\left(\frac{\partial \widetilde{B}(\vec{
\delta q},t)}{\partial \vec{\delta q_i}} \right)^2 ,
\label{eq:HV_tilde_ABB}
\end{equation}
is a function of only the coordinates $\vec{\xi}=(\vec{\delta q},t)$
and thus easily integrated
\begin{equation}
e^{\tau L_{\widetilde{C}}}: \left\{ \begin{array}{lll} \vec{\delta q}' & = &
\vec{\delta q}\\ t' & = & t\\ \vec{\delta p}' & = & \displaystyle \vec{\delta
p} - \frac{\partial \widetilde{C}( \vec{
\delta q},t)}{\partial \vec{\delta q}}  \tau \\ \delta p'_{N+1} & = & \displaystyle \delta p_{N+1} -
\frac{\partial \widetilde{C}(\vec{ \delta q},t)}{\partial t}  \tau
\end{array}\right. .
\label{eq:HV_tilde_LC}
\end{equation}
The explicit expressions of these maps for the H\'enon-Heiles system
(\ref{eq:HH}) are given in Appendix \ref{EPS_HH}.

From equations (\ref{eq:HV_tilde_LA}), (\ref{eq:HV_tilde_LB}) and
(\ref{eq:HV_tilde_LC}) we see that time $t$ is changed only by the act
of operator $e^{\tau L_{\widetilde{A}}}$. On the other hand, operators
$e^{\tau L_{\widetilde{B}}}$ and $e^{\tau L_{\widetilde{C}}}$ require
the knowledge of positions $\vec{q}$ at specific times for the
evaluation of the partial derivatives of $\widetilde{B} $ and
$\widetilde{C} $. We also note that for all these operators the last
equation for $\delta p_{N+1}$ can be neglected, since the knowledge of
its value does not influence the evolution of the other quantities,
and consequently the solution of the variational equations
(\ref{eq:w_V}).

Since the coordinates of the orbit $\vec{q}$ are known only at
specific times $t_i=i\Delta t$, $i=0,1,\ldots$, one is restricted to
use integration schemes that require the knowledge of $\vec{q}$ at
exactly these times. Such a scheme is, for example, the $S_{B1}$
integrator
\begin{equation}
S_{B1}= e^{(\tau/2) L_{\widetilde{B}}} e^{\tau L_{\widetilde{A}}}
e^{(\tau/2) L_{\widetilde{B}}}
\label{eq:SBAB1}
\end{equation}
(which is practically the well-known St\"{o}rmer/Verlet or leap-frog
method) with $\tau=\Delta t$. The right operator $e^{(\tau/2)
  L_{\widetilde{B}}}$ which acts first, requires the knowledge of
$\vec{q}(t_i)$, while the left operator $e^{(\tau/2)
  L_{\widetilde{B}}}$ needs the values of
$\vec{q}(t_i+\tau)=\vec{q}(t_{i+1})$, because the time value has
changed from $t_i$ to $t_i+\tau$ by $e^{\tau L_{\widetilde{A}}}$. Note
that the $S_{A1}$ integrator
\begin{equation}
S_{A1}= e^{(\tau/2) L_{\widetilde{A}}} e^{\tau L_{\widetilde{B}}}
e^{(\tau/2) L_{\widetilde{A}}}
\label{eq:SABA1}
\end{equation}
requires the knowledge of $\vec{q}(t_i+\tau/2)$ for the application of
$e^{\tau L_{\widetilde{B}}}$. This second order integration scheme
could be used with $\tau=2 \Delta t$, leading in general to a less
accurate algorithm compared to $S_{B1}$ (\ref{eq:SBAB1}), which is
also a second order integrator but uses a smaller time step $\tau=
\Delta t$. For $\tau=2 \Delta t$ is in general more efficient to apply the 
integration scheme
\begin{equation}
S_{B2}= e^{(\tau/6) L_{\widetilde{B}}} e^{(\tau/2) L_{\widetilde{A}}}
e^{(2\tau/3) L_{\widetilde{B}}} e^{(\tau/2) L_{\widetilde{A}}} e^{(\tau/6)
L_{\widetilde{B}}},
\label{eq:SBAB2}
\end{equation}
which was initially derived in \cite{ML95}. This integrator needs the
known values $\vec{q}(t_i)$, $\vec{q}(t_i+\tau/2)=\vec{q}(t_i+\Delta
t)=\vec{q}(t_{i+1})$ and $\vec{q}(t_i+\tau)=\vec{q}(t_i+2\Delta
t)=\vec{q}(t_{i+2})$.

The above
integration schemes can also be combined with a corrector step, since
$e^{\tau L_{\widetilde{C}}}$ (\ref{eq:HV_tilde_LC}) does not change
the time values, and acts before and after the main body of the
integrator (see equation (\ref{eq:saba2c})), when $t$ has values for
which we know the coordinates $\vec{q}$. We refer to this technique as
the \textit{TDHeps method} (eps: extended phase space). For the
numerical applications of the TDHeps method (presented in
Sect.~\ref{sec:numerics}) we use the fourth order integrator $S_{B2}^c$
both for the integration of the variational equations and for the
computation of the orbit.

Higher order SBAB or SABA integrators cannot be used in this
framework, because they require the knowledge of $\vec{q}$ at non
equidistant time values, different from $t_i$. In order to apply such
schemes one could initially compute the solution of equations
(\ref{eq:eq_mot_V}) also at these specific times (e.~g.~by
interpolation), but this would lead to a cumbersome, complex, time
consuming, and consequently inefficient scheme.

\subsection{The tangent map (TM) method}
\label{sec:TMM_gen}

The set of equations (\ref{eq:eq_mot_V}) and (\ref{eq:w_V}) can be
considered as a unified set of differential equations
\begin{equation}
\left.
\begin{array}{l}
\displaystyle \dot{\vec{q}}=\vec{p} \\
\displaystyle \dot{\vec{p}}=-\frac{\partial V(\vec{q})}{\partial \vec{q}} \\
\displaystyle  \dot{\vec{\delta q}}=\vec{\delta p} \\
\displaystyle
  \displaystyle \dot{\vec{\delta p}} =-
\mathbf{D}^2_V(\vec{q})\vec{\delta q}
\end{array} \right\} \Rightarrow \frac{d \vec{u}} {dt}= L_{HV}
\vec{u},
\label{eq:L_HV}
\end{equation}
where $ \vec{u} = (\vec{q},\vec{p}, \vec{\delta q}, \vec{\delta p})$
is a vector formed by the phase space vector
$\vec{x}=(\vec{q},\vec{p})$ and the deviation vector
$\vec{w}=(\vec{\delta q}, \vec{\delta p})$, and $L_{HV}$ is the
differential operator of the whole system. In analogy to equation
(\ref{eq:Ham_sol}), the solution of system (\ref{eq:L_HV}) for an
initial condition $\vec{u}(0)$ can be formally written as
$\vec{u}(t)=e^{t L_{HV}} \vec{u}(0)$. We describe now how symplectic
integrators can be used to obtain this solution.

First of all, let us note that equations (\ref{eq:L_HV}) cannot be
considered as the Hamilton's equations of motion of some generalized
Hamiltonian function. If such a Hamiltonian existed, and could be split
into two integrable parts, any symplectic integrator could be used for
finding the solution of system (\ref{eq:L_HV}). Since this is not the
case, we follow a different approach to achieve this goal. In section
\ref{sec:SI} the integration of the equations of motion of Hamiltonian
(\ref{eq:HH_usual}) over one integration time step $\tau$ was split
into steps over appropriate time intervals $c_i \tau$, $d_i \tau$,
where the dynamics was determined either by Hamiltonian $A(\vec{p})$
or $B(\vec{q})$ (\ref{eq:H_AB}). During these inter\-mediate steps the
tangent dynamics of the system is governed by the variational
equations
\begin{equation}
\begin{array}{l}
\displaystyle  \dot{\vec{\delta q}}=\vec{\delta p} \\
\displaystyle
  \displaystyle \dot{\vec{\delta p}} =0
\end{array}  \label{eq:varA_HV}
\end{equation}
for $A(\vec{p})$, and by
\begin{equation}
\begin{array}{l}
\displaystyle  \dot{\vec{\delta q}}=0 \\
\displaystyle
  \displaystyle \dot{\vec{\delta p}} =-
\mathbf{D}^2_V(\vec{q})\vec{\delta q}
\end{array} \label{eq:varB_HV}
\end{equation}
for $B(\vec{q})$.  Therefore, for each intermediate step of the
symplectic integration scheme the dynamics of the phase and the
tangent space is governed by the set of equations
\begin{equation}
\left.
\begin{array}{l}
\displaystyle \dot{\vec{q}}=\vec{p} \\
\displaystyle \dot{\vec{p}}=0 \\
\displaystyle  \dot{\vec{\delta q}}=\vec{\delta p} \\
\displaystyle
  \displaystyle \dot{\vec{\delta p}} =0
\end{array} \right\} \Rightarrow \frac{d \vec{u}} {dt}= L_{AV}
\vec{u}, \label{eq:L_AV}
\end{equation}
and
\begin{equation}
\left.
\begin{array}{l}
\displaystyle \dot{\vec{q}}=0 \\
\displaystyle \dot{\vec{p}}=-\frac{\partial V(\vec{q})}{\partial \vec{q}} \\
\displaystyle  \dot{\vec{\delta q}}=0 \\
\displaystyle
  \displaystyle \dot{\vec{\delta p}} =-
\mathbf{D}^2_V(\vec{q})\vec{\delta q}
\end{array} \right\} \Rightarrow \frac{d \vec{u}} {dt}= L_{BV}
\vec{u}, \label{eq:L_BV}
\end{equation}
for Hamiltonians $A(\vec{p})$ and $B(\vec{q})$ (\ref{eq:H_AB})
respectively, with $L_{AV}$ and $L_{BV}$ being the corresponding
differential operators.

These sets of equations are immediately solved, leading to maps
\begin{equation}
e^{\tau L_{AV}}: \left\{ \begin{array}{lll}
\vec{q}\,' & = &  \vec{ q} + \vec{p} \tau\\
\vec{ p}' & = & \vec{p} \\
\vec{\delta q}' & = &   \vec{\delta q} + \vec{\delta p} \tau \\
\vec{\delta p}' & = & \vec{\delta p}
\end{array}\right. ,
\label{eq:H_LAV}
\end{equation}
\begin{equation}
e^{\tau L_{BV}}: \left\{
\begin{array}{lll}
\vec{q}\,'  & = &  \vec{q}\\
\vec{p}\,' & = &  \displaystyle \vec{p} - \frac{\partial V(\vec{ q})}{\partial
\vec{ q}}   \tau  \\
\vec{\delta q}'  & = &  \vec{\delta q}\\
\vec{\delta p}' & = &  \displaystyle \vec{\delta p} -
\mathbf{D}^2_V(\vec{q})\vec{\delta q} \tau \end{array}
\right. . \label{eq:H_LBV}
\end{equation}
Obviously the first two equations of maps $e^{\tau L_{AV}}$, $e^{\tau
  L_{BV}}$ are exactly maps $e^{\tau L_{A}}$ (\ref{eq:H_LA}) and
$e^{\tau L_{B}}$ (\ref{eq:H_LB}), respectively.

Thus, \textbf{any symplectic integration scheme used to solve the
  Hamilton's equations of motion (\ref{eq:eq_mot_V}), which involves
  the successive application of maps $\mathbf{e^{\tau L_{A}}}$
  (\ref{eq:H_LA}), $\mathbf{e^{\tau L_{B}}}$ (\ref{eq:H_LB}), can also
  be used for the simultaneous integration of the variational
  equations (\ref{eq:w_V}), i.~e.~for solving the set of equations
  (\ref{eq:L_HV}), by replacing maps $\mathbf{e^{\tau L_{A}}}$,
  $\mathbf{e^{\tau L_{B}}}$ with maps $\mathbf{e^{\tau L_{AV}}}$
  (\ref{eq:H_LAV}) and $\mathbf{e^{\tau L_{BV}}}$ (\ref{eq:H_LBV})
  respectively.} This statement is a specific application of a more
general result which is stated for example in \cite{LR01}: Symplectic
integration schemes can be applied to first order differential systems
$\dot{X}=LX$ that can be written in the form $\dot{X}=(L_A+L_B)X$,
where $L$, $L_A$ and $L_B$ are differential operators for which the
two systems $\dot{X}=L_A X$ and $\dot{X}=L_B X$ are integrable. The
system of differential equations $\dot{u}=L_{HV} u$ (\ref{eq:L_HV})
belongs to this category since it can be split into the integrable
systems $\dot{u}=L_{AV} u$ (\ref{eq:L_AV}) and $\dot{u}=L_{BV} u$
(\ref{eq:L_BV}).

Let us discuss this splitting in more detail. The system
(\ref{eq:L_HV}) can be written as
\begin{equation}
\begin{array}{l}\displaystyle
  \dot{\vec{\mathcal{Q}}}=\vec{\mathcal{P}} \\ \displaystyle
  \dot{\vec{\mathcal{P}}} =\vec{\mathcal{F}}(\vec{\mathcal{Q}})
\end{array}
\label{eq:new_QP}
\end{equation}
with $ \vec{\mathcal{Q}}=(\vec{q},\vec{\delta
  q})=(q_1,q_2,\ldots,q_N,\delta q_1,\delta q_2, \ldots, \delta q_N)$,
$\vec{\mathcal{P}}=(\vec{p}, \vec{\delta
  p})=(p_1,p_2,\ldots,p_N,\delta p_1,\delta p_2, \ldots, \delta p_N)$,
and $\vec{\mathcal{F}}(\vec{\mathcal{Q}})$ being a vector with
coordinates
\begin{equation}
\mathcal{F}_i = \left\{ \begin{array}{lll}
\displaystyle -\frac{\partial V(\vec{q})}{\partial \vec{q}_i} & \mbox{for} & 1\leq i \leq N,\\ \\
\displaystyle - \sum_{k=1}^{N}\frac{\partial^2
      V(\vec{q})}{\partial q_i \partial q_k} \delta q_k\,
& \mbox{for} & N<i \leq 2N
\end{array}\right. .
\label{eq:F_Q}
\end{equation}
Then the dynamics of any general variable $U(\vec{\mathcal{Q}},\vec{\mathcal{P}})$ is given by
\begin{equation}
\begin{array}{lll}
  \displaystyle
  \dot{U}(\vec{\mathcal{Q}},\vec{\mathcal{P}})& =& \displaystyle \sum_{i=1}^{2N} \left[ \frac{\partial U(\vec{\mathcal{Q}},\vec{\mathcal{P}})}{\partial \mathcal{Q}_i}\dot{\mathcal{Q}_i} + \frac{\partial U(\vec{\mathcal{Q}},\vec{\mathcal{P}})}{\partial \mathcal{P}_i}\dot{\mathcal{P}_i} \right] \stackrel{\mbox{(\ref{eq:new_QP})}}{=}\\ \\
  &=& \displaystyle \left\lbrace \sum_{i=1}^{2N} \left[ \mathcal{P}_i\frac{\partial }{\partial \mathcal{Q}_i} + \mathcal{F}_i \frac{\partial }{\partial \mathcal{P}_i} \right] \right\rbrace U(\vec{\mathcal{Q}},\vec{\mathcal{P}})=\\ \\
  \displaystyle
  &=& \displaystyle \left( L_{AV}+L_{BV}\right)  U(\vec{\mathcal{Q}},\vec{\mathcal{P}}).
\end{array} 
\label{eq:evol_U}
\end{equation}
The solution of Eq.~(\ref{eq:evol_U}) for a time step $\tau$ can be
formally written as
\begin{equation}
U(t+\tau)=e^{\tau(L_{AV}+L_{BV})} U(t).
\label{eq:QP_tau}
\end{equation}
The decomposition of $e^{\tau(L_{AV}+L_{BV})} $ into products of
operators $e^{\tau L_{AV}} $, $e^{\tau L_{BV}} $ by any symplectic
integration scheme gives rise to an exponential-splitting algorithm
for the integration of system (\ref{eq:L_HV}), which would be
symplectic if Eqs.~(\ref{eq:L_HV}) were the equations of motion of a
Hamiltonian function (which are not, as we have already discussed).

In our study we consider symplectic integrators that require the
application of corrector terms. When the $S_{Bn}^c$ ($S_{An}^c$)
integrators are used, map $e^{\tau L_C}$ (\ref{eq:H_LC}) acts for some
intermediate steps of the algorithm.  Formally one can consider that
for these steps the phase space dynamics is governed by the Hamilton's
equations of motion of the Hamiltonian function $C(\vec{q})$
(\ref{eq:H_ABB}) (whose solution is given by map $e^{\tau L_{C}}$
(\ref{eq:H_LC})). Consequently, the tangent space dynamics is
described for these time steps by the variational equations of
Hamiltonian $C(\vec{q})$. So the evolution of the general vector
$\vec{u}$ is given by
\begin{equation}
\left.
\begin{array}{l}
\displaystyle \dot{\vec{q}}=0 \\
\displaystyle \dot{\vec{p}}=-\frac{\partial C(\vec{q})}{\partial \vec{q}} \\
\displaystyle  \dot{\vec{\delta q}}=0 \\
\displaystyle
  \displaystyle \dot{\vec{\delta p}} =-
\mathbf{D}^2_C(\vec{q})\vec{\delta q}
\end{array} \right\} \Rightarrow \frac{d \vec{u}} {dt}= L_{CV}
\vec{u}, \label{eq:L_CV}
\end{equation}
where $\textbf{D$^2_C$}(\vec{q})_{jk}= \partial^2 C(\vec{q}) / \partial
q_j \partial q_k$.  We easily see that the solution of these equations
is given by the map
\begin{equation}
e^{\tau L_{CV}}: \left\{
\begin{array}{lll}
\vec{q}\,'  & = &  \vec{q}\\
\vec{p}\,' & = &  \displaystyle \vec{p} - \frac{\partial C(\vec{q})}{\partial
\vec{ q}}   \tau  \\
\vec{\delta q}'  & = &  \vec{\delta q}\\
\vec{\delta p}' & = &  \displaystyle \vec{\delta p} -
\mathbf{D}^2_C(\vec{q})\vec{\delta q} \tau \end{array}
\right. , \label{eq:H_LCV}
\end{equation}
which, of course, is an extension of map $e^{\tau L_{C}}$
(\ref{eq:H_LC}). So \textbf{the use of the corrector term with the
$\mathbf{S_{Bn}}$ ($\mathbf{S_{An}}$) integrator for the integration
  of system (\ref{eq:L_HV}) requires the additional substitution of
  map $\mathbf{e^{\tau L_{C}}}$ (\ref{eq:H_LC}) by the extended map
  $\mathbf{e^{\tau L_{CV}}}$ (\ref{eq:H_LCV}).}

We call the above-described procedure for the simultaneous integration
of the Hamilton's equations of motion (\ref{eq:eq_mot_V}) and the
variational equations (\ref{eq:w_V}), the \textit{tangent map (TM)
  method}. The explicit expressions of the extended maps $e^{\tau
  L_{AV}}$ (\ref{eq:H_LAV}), $e^{\tau L_{BV}}$ (\ref{eq:H_LBV}) and
$e^{\tau L_{CV}}$ (\ref{eq:H_LCV}) for the H\'enon-Heiles system
(\ref{eq:HH}) are given in Appendix \ref{TMM_HH}.

\section{Numerical applications}
\label{sec:numerics}

In order to study the efficiency of the different schemes for the
integration of the variational equations, we apply them to some simple
Hamiltonian systems of different numbers of degrees of freedom. In
particular we consider a) the well-known 2D H\'enon-Heiles system
\cite{HH64} described by the Hamiltonian
\begin{equation}
H_2 = \frac{1}{2} (p_x^2+p_y^2) + \frac{1}{2} (x^2+y^2) + x^2 y -
\frac{1}{3} y^3,
\label{eq:HH}
\end{equation}
b)  the 3D Hamiltonian system
\begin{equation}
H_3 = \frac{1}{2} (x^2+p_x^2) + \frac{\sqrt{2}}{2} (y^2+p_y^2) +
\frac{\sqrt{3}}{2} (z^2+p_z^2) + x^2 y + x^2 z,
\label{eq:3DHam}
\end{equation}
studied in \cite{CGG78,BGGS80b,SBA07}, and c) the famous
Fermi-Pasta-Ulam (FPU) $\beta$-lattice model \cite{FPU}, which
describes a chain of $N$ particles with nearest neighbor interaction,
for the particular case of $N=8$ studied in \cite{SBA08}. The 8D
Hamiltonian of this system is
\begin{equation}
H_8 = \sum_{i=1}^{8} \frac{p_i^2}{2} + \sum_{i=0}^{8}
\left[\frac{(q_{i+1}-q_i)^2}{2} + \frac{\beta (q_{i+1}-q_i)^4}{4} \right] \, .
\label{eq:FPUHam}
\end{equation}

We consider some typical regular and chaotic orbits of these systems
and investigate the efficiency of the various numerical techniques by
checking how well their outcomes verify the following theoretically
known properties of the LCEs and the GALIs:
\begin{itemize}
\item The finite time mLCE $X_1(t)$ should eventually tend to zero in
  the case of regular orbits following the power law given in
  (\ref{eq:X1_t}).

\item According to Eq.~(\ref{eq:op_signs}), the LCEs are grouped in
  pairs of values having opposite signs, and consequently their sum
  vanishes. Therefore the same relation should be also satisfied by
  the limiting values of the corresponding finite time LCEs i.~e.
\begin{equation}
\lim_{t\rightarrow \infty} \left( X_i(t) + X_{2N-i+1}(t)\right) = 0 \,\,\, , \,\,\, i=1,2,\ldots, N.
\label{eq:op_signs_X}
\end{equation}

\item According to Eq.~(\ref{eq:spct_00}) at least two LCEs vanish and
  therefore $X_N(t)$ and $X_{N+1}(t)$ should tend to zero.

\item The GALIs follow the laws (\ref{eq:GALI_chaos}) and
  (\ref{eq:GALI_order_all_N}) for chaotic and regular orbits
  respectively.
\end{itemize}

\subsection{The  2D H\'enon-Heiles system}
\label{sec:HH}

We implement first the various numerical schemes presented in
Sect.~\ref{sec:num_int} for the integration of the variational
equations of regular and chaotic orbits of the 2D H\'enon-Heiles
system (\ref{eq:HH}). The explicit expressions of all these schemes
are presented in detail in Appendix \ref{ap:HH_var_eq_all}.  The
orbits of the H\'enon-Heiles system have four LCEs $\chi_1\geq
\chi_2\geq \chi_3\geq \chi_4$, with $\chi_2=\chi_3=0$ and
$\chi_1=-\chi_4\geq0$. A simple qualitative way of studying the
dynamics of a Hamiltonian system is to plot the successive
intersections of its orbits with a Poincar\'{e} surface of section
(PSS) (see for example Sect.~1.2b of \cite{LichtenbergL_1992}). In 2D
systems like (\ref{eq:HH}), the PSS is a two dimensional plane which
allows the clear visualization of the dynamics.

\begin{figure}
\includegraphics[scale=0.45]{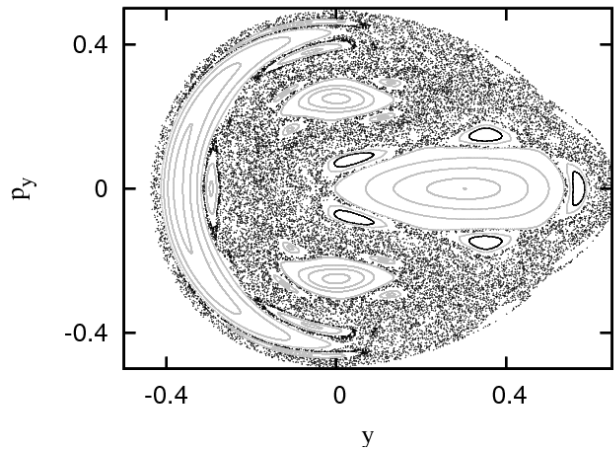}
\caption{ The PSS defined by $x=0$, $p_x \geq 0$, for the
  H\'enon-Heiles system (\ref{eq:HH}) with $H_2=0.125$. The regular
  orbit R1 corresponds to the five closed black curves around the
  right large island of stability, while the chaotic orbit C1 is
  represented by the black dots scattered over the PSS. In order to
  get a clear picture of the structure of the whole PSS, other orbits
  of the system are plotted in gray.}
\label{f:HHPSS}
\end{figure}

In our study we keep the value of the Hamiltonian fixed at
$H_2=0.125$.  Initially, we consider two representative orbits of the
system: the regular orbit R1 with initial conditions $x=0$,
$p_x\approx0.2334$, $y=0.558$, $p_y=0$, and the chaotic orbit C1 with
initial conditions $x=0$, $p_x\approx0.4208$, $y=-0.25$, $p_y=0$.  In
Fig.~\ref{f:HHPSS}
we plot the intersection points of these two orbits with the PSS
defined by $x=0$, $p_x\geq0$. The points of the regular orbit lie on a
torus and form five smooth closed curves (the so-called stability
islands) on the PSS, while the points of the chaotic orbit appear
randomly scattered.

First, we use the DOP853 non-symplectic scheme to integrate the set of
differential equations composed from the Hamilton's equations of motion
(\ref{eq:HHeq}) and the variational equations
(\ref{eq:HH_var_eq}). In our
computations we set the integration time
step $\tau=0.05$ and the threshold parameter $\delta=10^{-5}$, unless
otherwise stated.

We also implement the TDHcc, the TDHes and the TDHeps methods.  For
these methods we initially integrate equations (\ref{eq:HHeq}) by the
$S_{B2}^c$ scheme.  In this way we obtain the coordinates of the orbit
at times $t_i=i \Delta t$, $i=0,1,2,\ldots$, with $\Delta t$ being the
constant integration step. Then we assume the TDH (\ref{eq:HH_var}) to
have constant coefficients in each time interval $[t_i,t_i+\Delta t)$
and either we integrate in this interval its equations of motion by
the $S_{B2}^c$ integrator (\mbox{TDHcc} method), or we compute the
exact solution of these equations by performing the canonical
transformation induced by matrix $\mathbf{T}$ of Eq.~(\ref{eq:T_HH})
(\mbox{TDHes} method). Alternatively, we use the $S_{B2}^c$ scheme for
integrating the equations of motion of the 3D Hamiltonian
$\displaystyle \widetilde{H}_{VH}$ (\ref{eq:HH_var_EPS}) in the time
interval $[t_i,t_i+2\Delta t)$, by applying
Eqs.~(\ref{eq:HVH_tilde_LA}), (\ref{eq:HVH_tilde_LB}) and
(\ref{eq:HVH_tilde_LC}) with time step $\tau=2\Delta t$ (TDHeps
method).  Finally we implement the TM method using the $S_{B2}^c$
integrator, which requires the application of maps
(\ref{eq:HH_H_LAV}), (\ref{eq:HH_H_LBV}) and (\ref{eq:HH_H_LCV}).

As a final remark we note that in all the above-described schemes
after each time step $\tau$ the LCEs (GALIs) are computed and the
deviation vectors are orthonormalized (normalized) having norm equal
to 1.

\subsubsection{Regular orbits}
\label{sec:HH_reg}

\begin{figure*}
\includegraphics[angle=0,width=2.05\columnwidth]{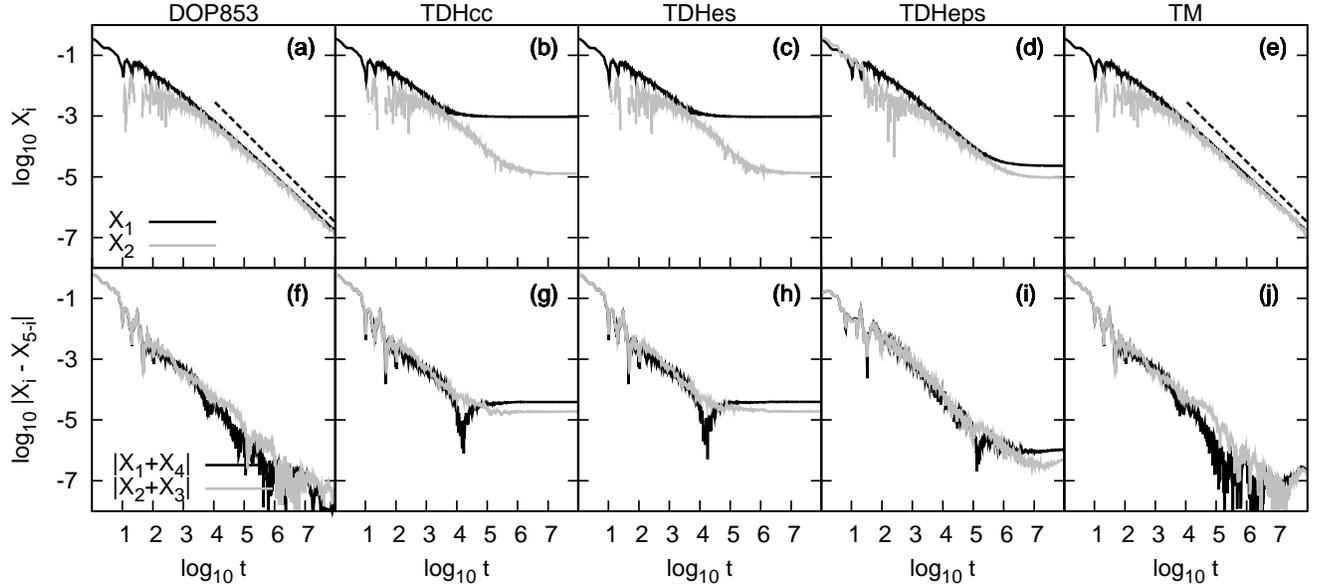}
\caption{The time evolution of $X_1(t)$ (black curves), $X_2(t)$ (gray
  curves) [upper panels] and $|X_1(t)+X_4(t)|$ (black curves),
  $|X_2(t)+X_3(t)|$ (gray curves) [lower panels] in log-log plots for
  the regular orbit R1 of the H\'enon-Heiles system (\ref{eq:HH}). The
  variational equations are integrated by the DOP853 integrator ((a)
  and (f)), and by the TDHcc ((b) and (g)), the TDHes ((c) and (h)),
  the TDHeps ((d) and (i)) and the TM ((e) and (j)) method. Dashed
  lines in panels (a) and (e) correspond to functions proportional to
  $t^{-1}$. The step
  size is $\tau=0.05$ for all methods. For the DOP853 method the
  parameter $\delta=10^{-5}$ is used. }
\label{f:HH_reg_var}
\end{figure*}

Results concerning the LCEs of the regular orbit R1 are shown in
Fig.~\ref{f:HH_reg_var}.
In particular, the time evolution of the finite time LCEs $X_1$ and
$X_2$ is given in the upper panels, while in the lower panels the
evolution of quantities $|X_1+X_4|$, $|X_2+X_3|$ is plotted.

In Table \ref{t:HH} information on the computation of the whole
spectrum of LCEs of the R1 orbit up to $t=10^8$ is reported.  The
relative energy error, which could be considered as an indicator of
the goodness of the integration procedure of orbit R1, increases with
time for the DOP853 method, while it fluctuates around a constant
value for all other methods. The values of this error and of $X_1$ at
the end of the integration are reported in the table. The CPU time
needed on an ordinary personal computer by each method for the
integration of the equations of motion and the variational equations,
as well as for the computation of the spectrum of LCEs is also given.

\begin{table*} [!ht]
\centering
\begin{tabular}{lccccr}
\toprule
Integrator   & Method             &   Step size  $\tau$      & Relative energy error & $X_1$                   &   CPU time  \\ \toprule
DOP853       & [$\delta= 10^{-5}$]&  5 $\times 10^{-2}$       & 7 $\times 10^{-10}$        & 1.6 $\times 10^{-7}$  &  8h 18m     \\
$S_{B2}^c$       & TDHcc              &  5 $\times 10^{-2}$      & 2 $\times 10^{-8}$         & 9.4 $\times 10^{-4}$  &  5h 48m      \\
$S_{B2}^c$       & TDHes              &  5 $\times 10^{-2}$      & 2 $\times 10^{-8}$         & 9.4 $\times 10^{-4}$  &  5h 36m      \\
$S_{B2}^c$       & TDHeps             &  5 $\times 10^{-2}$      & 2 $\times 10^{-8}$         & 2.3 $\times 10^{-5}$  &  6h 03m      \\
$S_{B2}^c$       & TM                 &  5 $\times 10^{-2}$      & 2 $\times 10^{-8}$         & 1.5 $\times 10^{-7}$  &  4h 40m      \\
\hline                                                                                                                         
DOP853       & [$\delta= 10^{-5}$]&  1 $\times 10^{-1}$       & 4 $\times 10^{-7}$         & 1.6 $\times 10^{-7}$  &  4h 11m      \\ 
DOP853       & [$\delta= 10^{-10}$]&  1 $\times 10^{-1}$       & 4 $\times 10^{-7}$         & 1.6 $\times 10^{-7}$  &  4h 12m      \\ 
DOP853       & [$\delta= 10^{-5}$]&  2 $\times 10^{-1}$       & 2 $\times 10^{-4}$         & 2.4 $\times 10^{-7}$  &  2h 06m      \\ 
DOP853       & [$\delta= 10^{-10}$]&  2 $\times 10^{-1}$       & 2 $\times 10^{-4}$         & 2.5 $\times 10^{-7}$  &  2h 03m      \\ 
DOP853       & [$\delta= 10^{-5}$]&  5 $\times 10^{-1}$       & 8 $\times 10^{-1}$         & 1.1 $\times 10^{-6}$  &   50m      \\
DOP853       & [$\delta= 10^{-10}$]&  5 $\times 10^{-1}$       & 6 $\times 10^{-4}$         & -7.7 $\times 10^{-8}$  &  1h 40m      \\
$S_{B2}^c$       & TDHeps             &  1 $\times 10^{-1}$      & 1 $\times 10^{-6}$         & 8.9 $\times 10^{-5}$  &  3h 01m       \\ 
$S_{B2}^c$       & TDHeps             &  2 $\times 10^{-1}$      & 2 $\times 10^{-5}$         & 3.5 $\times 10^{-4}$  &  1h 33m       \\
$S_{B2}^c$       & TDHeps             &  5 $\times 10^{-1}$      & 1 $\times 10^{-3}$          & 1.8 $\times 10^{-3}$ &  37m          \\
$S_{B2}^c$       & TM                 &  1 $\times 10^{-1}$      & 2 $\times 10^{-6}$         & 1.6 $\times 10^{-7}$  &  2h 16m       \\ 
$S_{B2}^c$       & TM                 &  2 $\times 10^{-1}$      & 2 $\times 10^{-5}$         & 3.3 $\times 10^{-8}$  &  1h 08m       \\ 
$S_{B2}^c$       & TM                 &  5 $\times 10^{-1}$      & 1 $\times 10^{-3}$         & 5.4 $\times 10^{-8}$  &  27m          \\ 
\toprule
\end{tabular}
\caption{Information for the  computation  of the whole spectrum of LCEs for the regular orbit R1 of the H\'enon-Heiles system (\ref{eq:HH}), up to $t=10^8$. The non-symplectic  DOP853 algorithm and the symplectic $S_{B2}^c$  integrator are used. In the latter case the $S_{B2}^c$ scheme is used for the evolution of the orbit, while  different approaches are applied for the integration of the variational equations. Step size $\tau$ is the time between two successive evaluations of the LCEs. For the TDHcc, the TDHes and the TM methods, $\tau$ coincides with the integration time step $\Delta t$ of the orbit, while for the TDHeps method $\tau=2 \Delta t$. In the case of the DOP853 algorithm the integration over time $\tau$ is performed with a variable integration step, so that the local one-step error is kept smaller than $\delta$. The relative energy error and the estimated value $X_1$ of the mLCE at $t=10^8$ are given. The required CPU time for the implementation of each method on an ordinary personal computer (AMD Athlon 1GHz) is given in the last column. The first 5 cases (above the horizontal line) are the ones presented in Fig.~\ref{f:HH_reg_var}.}
\label{t:HH}
\end{table*}

The results of Fig.~\ref{f:HH_reg_var} show that the DOP853
(Fig.~\ref{f:HH_reg_var}(a)) and the TM method
(Fig.~\ref{f:HH_reg_var}(e)) have the best performance in evaluating
the mLCE, because $X_1$ tends to zero until the end time $t=10^8$ of
the integration, following a $t^{-1}$ law.  The good behavior of the
DOP853 and the TM methods is due to the fact that the first technique
is used for the integration of the actual set of Eqs.~(\ref{eq:HHeq})
and (\ref{eq:HH_var_eq}) which govern the dynamics of the orbit and
the deviation vector, while the second method approximates very
accurately the dynamics of the system by the repeated application of a
symplectic map, and the tangent dynamics by the act of the
corresponding tangent map.

For the TDHcc (Fig.~\ref{f:HH_reg_var}(b)), the TDHes
(Fig.~\ref{f:HH_reg_var}(c)) and the TDHeps
(Fig.~\ref{f:HH_reg_var}(d)) methods $X_1$ initially decreases too as
$X_1\propto t^{-1}$, but later its value deviates from the approximate
$t^{-1}$ law and tends to a constant (different for each method)
nonzero value.  Among these techniques the TDHeps method has the best
performance, because the computed $X_1$ levels off to smaller values
than in the cases of TDHcc and TDHes methods, being $X_1\approx 2.3
\times 10^{-5}$ at $t=10^8$. Nevertheless, from the results of
Figs.~\ref{f:HH_reg_var}(b)-(d) one would wrongly characterize the
regular orbit R1 as chaotic. Concerning the TDHcc and TDHes methods,
the main reason for this discrepancy is that these methods approximate
the tangent dynamics by considering constants the actual time
dependent coefficients of Hamiltonian $H_{VH}$ (\ref{eq:HH_var}), for
the duration of each integration time step.  The equations of motion
of $H_{VH}$ with constant coefficients are solved explicitly by the
TDHes method, while their solution is approximated by the application
of the TDHcc scheme. For the used time step $\tau=0.05$, both methods
give practically the same $X_1$ at $t=10^8$. For smaller time steps
the final values of $X_1$ obtained by both techniques are closer to
the theoretical value $X_1=0$. On the other hand, since the TDHeps
method takes into account the time dependent nature of the
coefficients of $H_{VH}$, it succeeds in obtaining a better estimation
of the mLCE compared to the TDHcc and the TDHes methods.

The computed values of the second largest LCE ($\chi_2=0$) have
similar characteristics with the results for the mLCE. Again, the
finite time LCE $X_2$ computed by the DOP853 integrator
(Fig.~\ref{f:HH_reg_var}(a)) and the TM method
(Fig.~\ref{f:HH_reg_var}(e)) tends to zero until the end of the
integration time. On the other hand, the $X_2$ computed by the TDHcc
(Fig.~\ref{f:HH_reg_var}(b)), the TDHes (Fig.~\ref{f:HH_reg_var}(c))
and the TDHeps (Fig.~\ref{f:HH_reg_var}(d)) methods does not tend to
zero, but levels off to positive values which are always smaller than
the level off values of $X_1$.  Again the TDHeps approach is more
accurate, because the final value $X_2\approx 9.3\times 10^{-6}$ at
$t=10^8$ obtained by this method is slightly smaller than the ones
found by the TDHcc and the TDHes methods, and thus closer to the real
$\chi_2=0$ value.

The ability of the DOP853 and the TM methods to evaluate quite
accurately the LCEs of the regular orbit R1 is also shown by the
tendency of quantities $|X_1+X_4|$, $|X_2+X_3|$ to become zero
(Fig.~\ref{f:HH_reg_var}(f) and (j)). Actually these quantities
attain, for both methods, very small values $\lesssim 10^{-7}$ at
$t=10^8$.  But when these quantities are computed by the other three
techniques they do not become zero as they theoretically should do,
but level off to small positive values
(Figs.~\ref{f:HH_reg_var}(g)-(i)).  Again the TDHeps method exhibits a
better performance since the level off values are smaller than the
ones obtained by the TDHcc and the TDHes methods.

Looking in Table \ref{t:HH} at the CPU times needed for the
computation of the whole spectrum of LCEs, one sees that the
non-symplectic method is  the most expensive one. Amongst the
remaining approaches the TM method is the fastest, due to the fact
that the whole set of equations for the evolution of both the orbit
and the deviation vector are integrated together. The TDHcc and TDHes
methods require more CPU time than the TM method, because for each
integration time step the evolutions of the orbit and the deviation
vectors are not performed simultaneously. First the orbit is
evolved. Its coordinates define the coefficients of $H_{VH}$
(\ref{eq:HH_var}), which are considered to be constant for the
duration of the time step. Then, the deviation vectors are advanced
for this particular Hamiltonian function for one time step. The TDHeps
method needs even more CPU time mainly because the orbit is integrated
with half time step ($\Delta t=\tau/2$) with respect to the other
methods.

\begin{figure}
\begin{tabular}{cc}
  \includegraphics[scale=1.15]{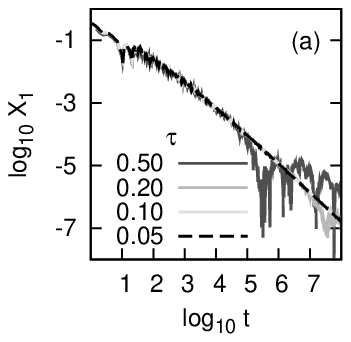} & \includegraphics[scale=1.15]{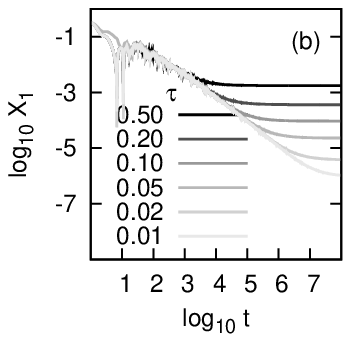}
\end{tabular}
\caption{The time evolution of $X_1(t)$ in log-log plots for the
  regular orbit R1 of the H\'enon-Heiles system (\ref{eq:HH}) for (a)
  the DOP853 (with $\delta=10^{-5}$), and (b) the TDHeps methods, when
  different step sizes $\tau$ are used. In (a) the curves for
  $\tau=0.05$ and $\tau=0.1$ practically overlap.}
\label{f:dif_steps}
\end{figure}

The first five rows of Table \ref{t:HH} contain information for the
particular cases shown in Fig.~\ref{f:HH_reg_var}. From these data we
see that the energy error for the DOP853 method at $t=10^8$, is
smaller than the error of the $S_{B2}^c$ integrator used by the other
methods. As it is also shown in Fig.~\ref{f:HH_reg_var} the values of
$X_1$ obtained by the DOP853 and the TM methods are close to each
other, despite the fact that the DOP853 method integrates orbit R1
with a better accuracy. Of major practical importance is the fact that
the DOP853 method needs almost two times more CPU time than the TM
method in order to compute the four LCEs up to $t=10^8$.  Increasing
the integration step size of DOP853 to $\tau=0.1$
(Fig.~\ref{f:dif_steps}(a)) still permits the computation of the same
$X_1$ value at $t=10^8$, but with a larger error in the conservation
of $H_2$. The $X_1$ computed by the DOP853 method for even larger step
sizes, like $\tau=0.2$ and $\tau=0.5$, starts after some time to
exhibit deviations from the $X_1 \propto t^{-1}$ law
(Fig.~\ref{f:dif_steps}(a)),
leading to somewhat larger final values ($X_1 \approx 2.4 \times
10^{-7}$ for $\tau=0.2$ and $X_1 \approx 1.1 \times 10^{-6}$ for
$\tau=0.5$) with respect to the $X_1 \approx 1.6 \times 10^{-7}$ value
found for smaller $\tau$. From our numerical experiments we see that
the required CPU time for the DOP853 method, as well as the relative
error of the computed energy $H_2$ mainly depend on the integration
time step $\tau$ and not on the threshold parameter $\delta$. In
particular, for $\tau \lesssim 0.2$ the value of $\delta$ does not
practically influence the required CPU time. For larger values of
$\tau$ (for which nevertheless the obtained results are not very
accurate) the CPU time is increased and the accuracy is improved when
$\delta$ is decreased. On the other hand, the TM method succeeds even
for $\tau=0.5$ to compute very fast the correct small final value of
$X_1 \lesssim 10^{-7}$. This method keeps also the relative energy
error at an acceptably low level, which is not the case any more for
the DOP853 method with the same time step.  Besides the computation
speed, this is an additional advantage of the TM method over the
DOP853 scheme.

It is worth noting that, although the DOP853 algorithm is an
integration scheme of higher order than the $S_{B2}^c$ symplectic
integrator used in the TM method, it shows worse characteristics than
the TM method, not only for large $\tau$, but also when we compare
implementations of the two algorithms that require almost the same CPU
time. For example, the DOP853 method for $\tau=0.2$ and
$\delta=10^{-10}$ (or even $\delta=10^{-5}$) has a final relative
energy error which is larger by 2 orders of magnitude with respect to
the error of the TM method for $\tau=0.1$ (which requires almost the
same CPU time $\approx 2$h, as seen in Table~\ref{t:HH}), and additionally the computed $X_1$
deviate from the $X_1 \propto t^{-1}$ law (Fig.~\ref{f:dif_steps}(a)).

Among the other applied methods which wrongly characterize the R1
orbit as chaotic, the TDHeps scheme has the best performance, since
$X_1$ eventually levels off to a small positive value. From the
results of Fig.~\ref{f:dif_steps}(b) we see that the decrease of the
step size $\tau$ pushes the starting time of the level off to larger
values and decreases the final value of $X_1$. So as one should
expect, smaller integration steps result in a more accurate
description of the evolution of the orbit and deviation vectors, and
leads to more accurate estimations of the LCEs. Nevertheless, the TM
method is preferred over the TDHeps method because for the same step
size $\tau$ it needs less CPU time, and  additionally it estimates more
accurately the LCEs.

For a regular orbit of the 2D Hamiltonian (\ref{eq:HH}) and a random
choice of initial deviation vectors, the theoretical prediction
(\ref{eq:GALI_order_all_N}) for the behavior of the GALIs gives
\begin{equation}
G_2(t) \propto \mbox{const.},\,\,\, G_3(t)
\propto \frac{1}{t^2},\,\,\,G_4(t) \propto
\frac{1}{t^4}. \label{eq:2DHam_or_GALIs}
\end{equation}

\begin{figure*}
\includegraphics[angle=0,width=2.05\columnwidth]{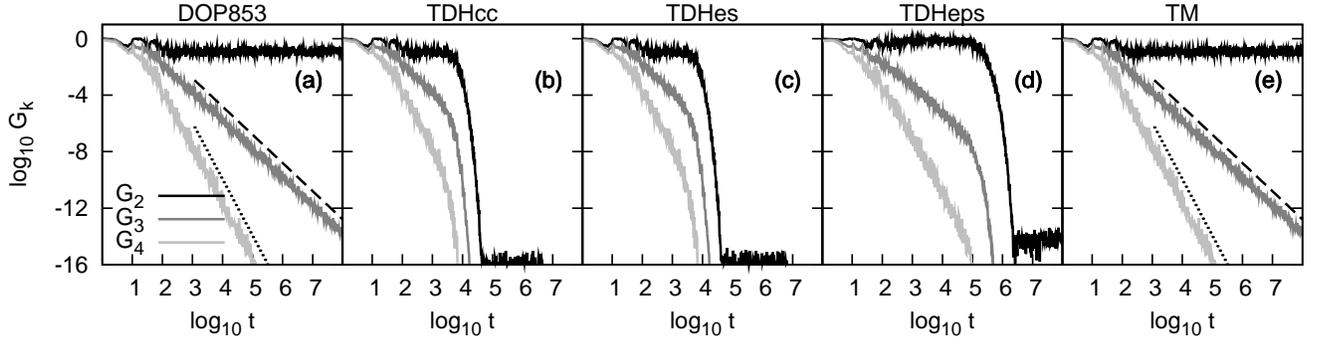}
\caption{The time evolution of $G_2(t)$ (black curves), $G_3(t)$
  (gray curves) and $G_4(t)$ (light gray curves) for the regular
  orbit R1 of the H\'enon-Heiles system (\ref{eq:HH}). The variational
  equations are integrated by the DOP853 (a), the TDHcc (b), the TDHes
  (c), the TDHeps (d), and the TM (e) method.  The plotted lines in
  panels (a) and (e) correspond to functions proportional to $t^{-2}$
  (dashed lines) and $t^{-4}$ (dotted lines). The values of $\tau$ and
  $\delta$ used in the integrations are the same as in
  Fig.~\ref{f:HH_reg_var}.}
\label{f:HH_reg_GALI}
\end{figure*}

In Fig.~\ref{f:HH_reg_GALI} we plot the time evolution of $G_2$,
$G_3$ and $G_4$ for the regular orbit R1, when the variational
equations are integrated by the same five numerical schemes used in
Fig.~\ref{f:HH_reg_var}.
The results obtained by the DOP853 (Fig.~\ref{f:HH_reg_GALI}(a)), and
the TM (Fig.~\ref{f:HH_reg_GALI}(e)) schemes are in accordance with
the theoretical predictions (\ref{eq:2DHam_or_GALIs}).  The GALIs
computed by the TDHcc (Fig.~\ref{f:HH_reg_GALI}(b)), the TDHes
(Fig.~\ref{f:HH_reg_GALI}(c)) and the TDHeps
(Fig.~\ref{f:HH_reg_GALI}(d)) methods follow the theoretical laws
(\ref{eq:2DHam_or_GALIs}) up to $t\approx10^4$ for the first two
methods and up to $t\approx10^5$ for the last one. After that time the
GALIs fall exponentially fast to zero indicating, wrongly, that the
orbit is chaotic. This behavior is in agreement with the behavior of
$X_1$ obtained by these methods in Fig.~\ref{f:HH_reg_var}, because
the mLCE levels off to a positive value after some initial time
interval, implying that the orbit is chaotic. The TDHeps method has
again a better performance than the other two methods used to
approximate the dynamics of the TDH (\ref{eq:HH_var}), since the
computed GALIs follow the theoretical predictions
(\ref{eq:2DHam_or_GALIs}) for longer times, but eventually it also
fails to characterize correctly the nature of orbit R1.

\subsubsection{Chaotic orbits}
\label{sec:HH_chaos}

\begin{figure*}
\includegraphics[angle=0,width=1.9\columnwidth]{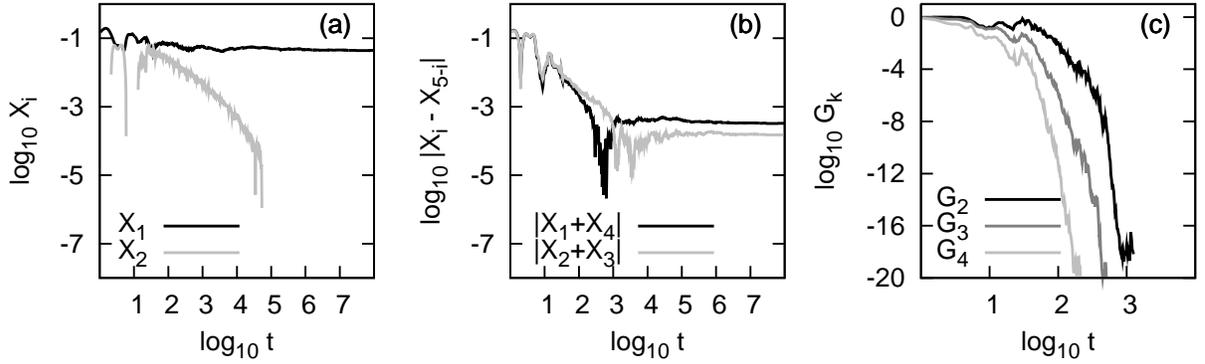}
\caption{The time evolution of (a) $X_1(t)$ (black curve), $X_2(t)$
  (gray curve), (b) $|X_1(t)+X_4(t)|$ (black curve),
  $|X_2(t)+X_3(t)|$ (gray curve), and (c)  $G_2(t)$ (black curve),
  $G_3(t)$ (gray curve) and $G_4(t)$ (light gray curve) for
  the chaotic orbit C1 of the H\'enon-Heiles system (\ref{eq:HH}) when the 
  variational equations are integrated by the DOP853 integrator. The
  values of $\tau$ and $\delta$ used in the integrations are the
  same as in Fig.~\ref{f:HH_reg_var}.}
\label{f:C1_LCES}
\end{figure*}

The computed LCEs and GALIs of the chaotic orbit C1 are practically
the same irrespectively of which of the previously presented methods
is used for the integration of the variational equations. For this
reason in Fig.~\ref{f:C1_LCES}
we present results obtained only by the DOP853 integrator.

From the results of Fig.~\ref{f:C1_LCES}(a) we see that $X_1$ remains
almost constant and different from zero, having practically the same
value $X_1\approx 4.5\times 10^{-2}$ at $t=10^{8}$ for all applied
schemes. Thus, all used methods are able to determine correctly the
chaotic nature of the orbit.  Since the H\'enon-Heiles system
(\ref{eq:HH}) is conservative, $\chi_2=0$. From
Fig.~\ref{f:C1_LCES}(a) we see that the finite time LCE $X_2$ tends to
zero, and becomes negative after $t\approx10^5$ with
$|X_2|<10^{-5}$. At that time all the applied numerical approaches
reach their limits of applicability for the accurate computation of
$\chi_2$.  The quantities $|X_1+X_4|$, $|X_2+X_3|$
(Fig.~\ref{f:C1_LCES}(b)), which theoretically should be zero, level
off after $t\approx10^3-10^4$ to $|X_1+X_4|\approx 4 \times 10^{-4}$
and $|X_2+X_3|\approx 10^{-4}$ for all used schemes. This behavior
indicates that all numerical methods succeed to reveal the symmetric
nature of the spectrum of LCEs but only up to four decimal digits of
accuracy. Finally, the computed values of GALIs of orbit C1
(Fig.~\ref{f:C1_LCES}(c)) show an exponential decay to zero which is a
characteristic of chaoticity.

\begin{figure}
\includegraphics[scale=1.4]{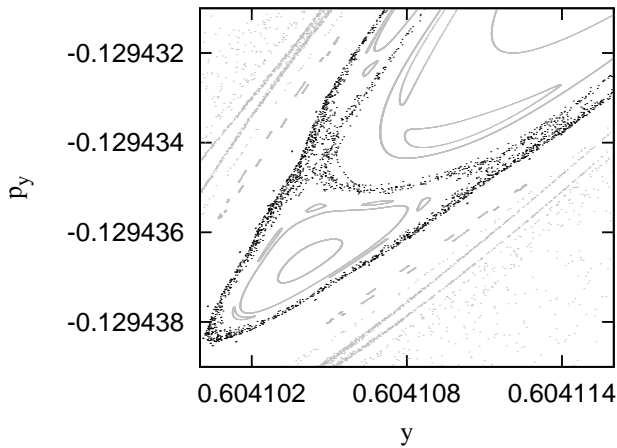}
\caption{ A part of the PSS ($x=0$, $p_x \geq 0$) of the
  H\'enon-Heiles system (\ref{eq:HH}) with $H_2=0.125$, where the
  weakly chaotic orbit C2 is plotted by black dots. }
\label{f:PSS_C2}
\end{figure}

Fig.~\ref{f:C1_LCES} shows the equivalence of the different numerical
techniques in the case of the chaotic orbit C1. This is a clear
difference with respect to the behavior of the various numerical
schemes for the regular orbit R1, where only the DOP853 and the TM
methods gave similar (to each other) and correct results
(Figs.~\ref{f:HH_reg_var} and \ref{f:HH_reg_GALI}). In order to check
if the equivalence of all methods is valid for all chaotic orbits we
consider a weakly chaotic orbit confined to a thin region of the phase
space at the borders of a small stability island
(Fig.~\ref{f:PSS_C2}).
We call this orbit C2 and its initial conditions are $x=0$,
$p_x\approx0.11879$, $y=0.335036$, $p_y=-0.385631$.

\begin{figure*}
\includegraphics[angle=0,width=2.05\columnwidth]{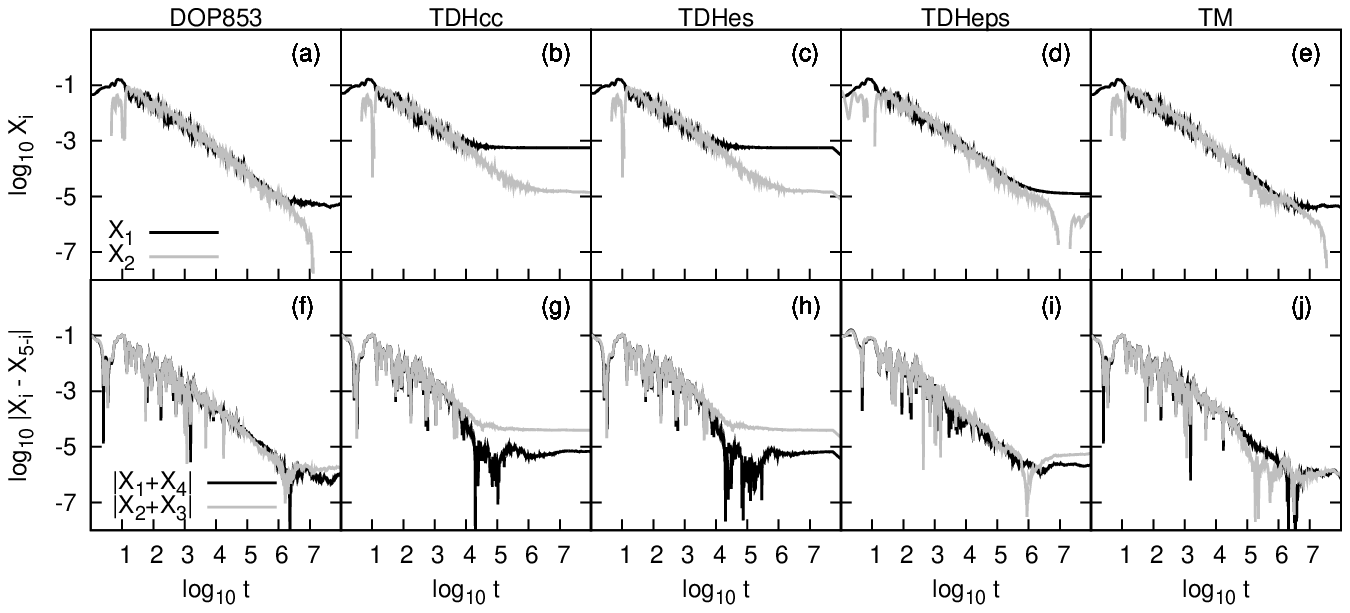}
\caption{The time evolution of $X_1(t)$ (black curves), $X_2(t)$
  (gray curves) [upper panels] and $|X_1(t)+X_4(t)|$ (black curves),
  $|X_2(t)+X_3(t)|$ (gray curves) [lower panels] in log-log plots for
  the chaotic orbit C2 of the H\'enon-Heiles system (\ref{eq:HH}). The
  variational equations are integrated by the DOP853 integrator ((a)
  and (f)), and by the TDHcc ((b) and (g)), the TDHes ((c) and (h)),
  the TDHeps ((d) and (i)) and the TM ((e) and (j)) method. The
  values of $\tau$ and $\delta$ used in the integrations are the
  same as in Fig.~\ref{f:HH_reg_var}. }
\label{f:C2_LCES}
\end{figure*}

From the results of the finite time LCEs of orbit C2 presented in
Fig.~\ref{f:C2_LCES},
we see that both the DOP853 (Fig.~\ref{f:C2_LCES}(a)) and the TM
(Fig.~\ref{f:C2_LCES}(e)) methods characterize orbit C2 as weakly
chaotic having a small mLCE $\chi_1\approx4 \times 10^{-6}$. The TDHcc
(Fig.~\ref{f:C2_LCES}(b)), the TDHes (Fig.~\ref{f:C2_LCES}(c)) and the
TDHeps (Fig.~\ref{f:C2_LCES}(d)) also characterize orbit C2 as chaotic
but overestimate the value of $\chi_1$. Thus, these three methods fail
to compute accurately the small value of the mLCE, with the TDHeps
method showing once more the best performance, because the computed
value ($X_1\approx 1.3\times 10^{-5}$) is closer to the real value of
$\chi_1$. The limitations of these three methods are also clearly seen
from the fact that the quantities $|X_1+X_4|$, $|X_2+X_3|$
(Figs.~\ref{f:C2_LCES}(g)-(i)) level off to larger values with respect
to the results obtained by the DOP853 (Fig.~\ref{f:C2_LCES}(f)) and
the TM (Fig.~\ref{f:C2_LCES}(j)) method. It is worth noting that the
level off values of $|X_1+X_4|$, $|X_2+X_3|$ obtained for orbit C2 by
the DOP853 and the TM methods are smaller than the saturation values of
the same quantities for the C1 orbit (Fig.~\ref{f:C1_LCES}(b)).

The results of Figs.~\ref{f:C1_LCES} and \ref{f:C2_LCES} lead us to
conclude that the DOP853 and the TM methods are able to accurately
compute mLCEs for a larger range of $\chi_1$ values than the TDHcc,
the TDHes and the TDHeps techniques. More specifically, our results
show that the DOP853 and the TM schemes can evaluate $\chi_1$ having
values at least as small as $10^{-6}$, while these small values
definitely exceed the computational ability of the TDHeps method
(which is the one with the best performance among the three other used
methods) for the used step size $\tau$.

\begin{figure*}
\includegraphics[angle=0,width=2.05\columnwidth]{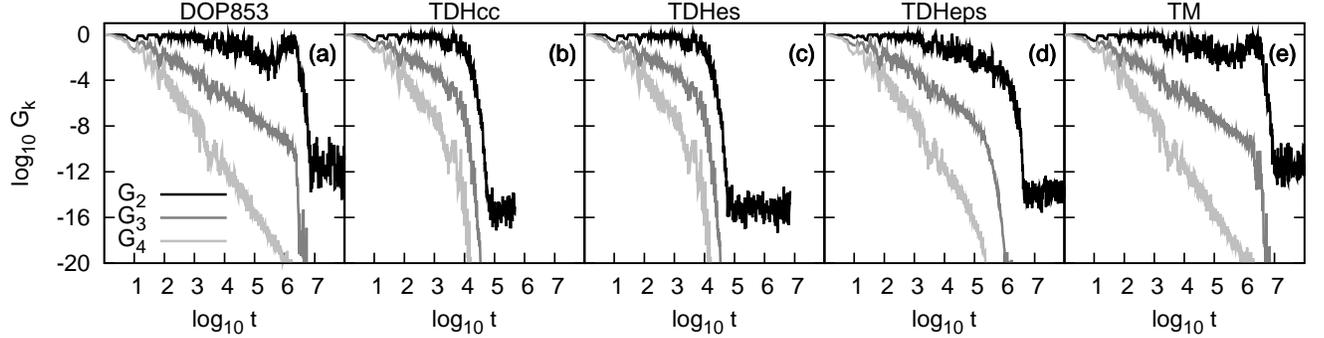}
\caption{The time evolution of $G_2(t)$ (black curves),
  $G_3(t)$ (gray curves) and $G_4(t)$ (light gray curves) for
  the chaotic orbit C2 of the H\'enon-Heiles system (\ref{eq:HH}). The
  variational equations are integrated by the DOP853 (a), the TDHcc
  (b), the TDHes (c), the TDHeps (d), and the TM (e) method. The
  values of $\tau$ and $\delta$ used in the integrations are the
  same as in Fig.~\ref{f:HH_reg_var}.}
\label{f:C2_GALIS}
\end{figure*}

The $G_k$, $k=2,3,4$ computed by the DOP853
(Fig.~\ref{f:C2_GALIS}(a))
and the TM (Fig.~\ref{f:C2_GALIS}(e)) method have practically the same
behavior. Up to $t\approx 10^6$, when the values of $X_1$ in
Figs.~\ref{f:C2_LCES}(a) and (e) start to level off deviating from the
$X_1 \propto t^{-1}$ law, the GALIs follow the theoretical predictions
(\ref{eq:2DHam_or_GALIs}) of regular motion. Later on the chaotic
behavior of orbit C2 becomes prominent and the GALIs fall
exponentially to zero. The time evolution of GALIs computed by the
TDHcc (Fig.~\ref{f:C2_GALIS}(b)), the TDHes (Fig.~\ref{f:C2_GALIS}(c))
and the TDHeps (Fig.~\ref{f:C2_GALIS}(d)) method also indicate that
the orbit is chaotic, but the exponential decay to zero starts
earlier.  This behavior is in accordance with the overestimation of
orbit's chaoticity, which was also seen in the computation of $X_1$
(Figs.~\ref{f:C2_LCES}(b)-(d)).

\subsection{Hamiltonian systems with more than two degrees of freedom}
\label{sec:multi_Ham}

Let us now apply the five different methods used in Sect.~\ref{sec:HH}
to regular and chaotic orbits of the 3D and the 8D Hamiltonian systems
(\ref{eq:3DHam}) and (\ref{eq:FPUHam}). In all studied cases the
computed LCEs and the GALIs have similar characteristics to the ones
seen for the 2D system (\ref{eq:HH}).  Due to the fact that the TM,
the DOP853 and the TDHeps methods always exhibited the best numerical
performance, we present in this section results obtained only by these
methods for the case of regular orbits.

\begin{figure}
\includegraphics[width=\columnwidth]{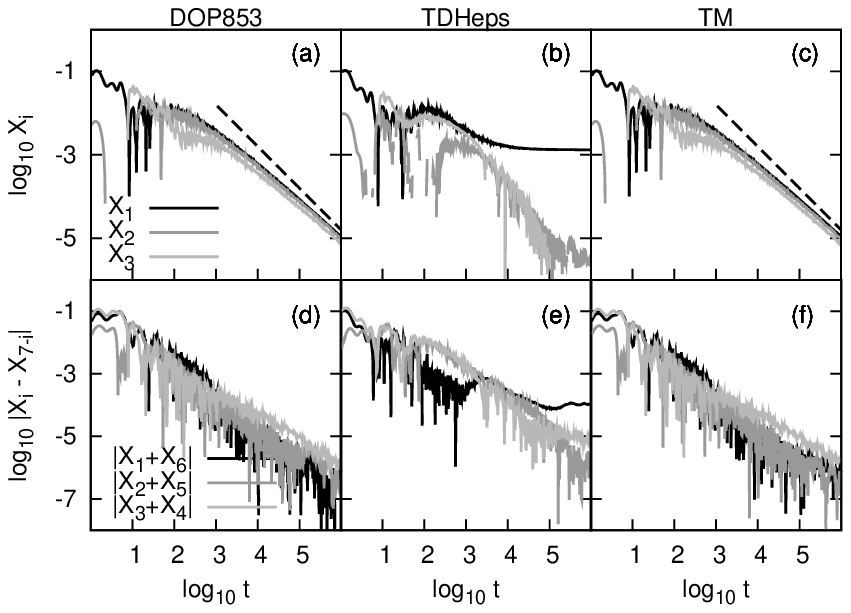}
\caption{The time evolution of $X_1(t)$, $X_2(t)$, $X_3(t)$ (upper
  panels) and $|X_1(t)+X_6(t)|$, $|X_2(t)+X_5(t)|$, $|X_3(t)+X_4(t)|$
  (lower panels) in log-log plots for the regular orbit R2 of the 3D
  Hamiltonian system (\ref{eq:3DHam}). The variational equations are
  integrated by the DOP853 integrator ((a) and (d)), and by the TDHeps
  ((b) and (e)) and the TM ((c) and (f)) method.  The step size is
  $\tau=0.05$ for all methods. For the DOP853 method the parameter
  $\delta=10^{-5}$ is used. Dashed lines in panels (a) and (c)
  correspond to functions proportional to $t^{-1}$. }
\label{f:3d_reg_lces}
\end{figure}

In Fig.~\ref{f:3d_reg_lces}
we show results for the six LCEs of a regular orbit with initial
conditions $x=y=z=0$, $p_x=0.1$, $p_y=0.347$, $p_z=0$ (orbit R2) of
the 3D system (\ref{eq:3DHam}), which was also studied in
\cite{SBA07}. Similarly to the results obtained for the 2D regular
orbit R1 in Fig.~\ref{f:HH_reg_var}, the three largest finite time
LCEs $X_1$, $X_2$, $X_3$ computed by the DOP853
(Fig.~\ref{f:3d_reg_lces}(a)) and the TM (Fig.~\ref{f:3d_reg_lces}(c))
method, tend to zero following a $X_i \propto t^{-1}$ $i=1,2,3$, law,
which indicates the regular nature of the orbit. These two methods are
also able to determine the symmetric nature of the spectrum of LCEs,
since the quantities $|X_1(t)+X_6(t)|$, $|X_2(t)+X_5(t)|$ and
$|X_3(t)+X_4(t)|$ tend to zero (Fig.~\ref{f:3d_reg_lces}(d) and
(f)). On the other hand, using the TDHeps method one would again
wrongly characterize the orbit as chaotic because the computed $X_1$
levels off at $t\approx10^4$ to a positive value, being $X_1 \approx
1.3 \times 10^{-3}$ at $t=10^6$ (Fig.~\ref{f:3d_reg_lces}(b)).  $X_2$
and $X_3$ show a better convergence to zero, while the latter one
becomes negative after $t\approx 10^5$ with $|X_3| < 10^{-5}$. In
addition, the quantity $|X_1(t)+X_6(t)|$ levels off to some finite
value, while $|X_2(t)+X_5(t)|$ and $|X_3(t)+X_4(t)|$ continue to
approach zero until the end of the integration
(Fig.~\ref{f:3d_reg_lces}(e)).

\begin{figure}
\includegraphics[angle=0,width=1.06\columnwidth]{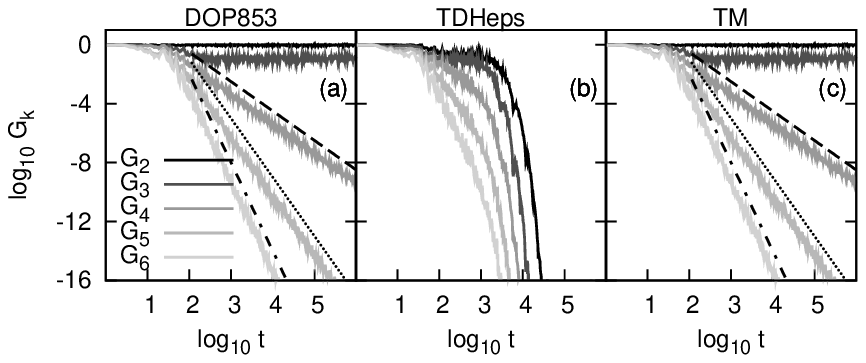}
\caption{The time evolution of $G_k(t)$, $k=2,3,\ldots,6$ for the
  regular orbit R2 of the 3D Hamiltonian system (\ref{eq:3DHam}).  The
  variational equations are integrated by the DOP853 (a), the TDHeps
  (b), and the TM (c) method. The values of $\tau$ and $\delta$ used
  in the integrations are the same as in Fig.~\ref{f:3d_reg_lces}. The
  plotted lines in panels (a) and (c) correspond to functions
  proportional to $t^{-2}$ (dashed line), $t^{-4}$ (dotted line) and
  $t^{-6}$ (dash-dotted line). }
\label{f:3d_reg_galis}
\end{figure}

According to Eq.~(\ref{eq:GALI_order_all_N}) the GALIs of a regular
orbit of the 3D Hamiltonian system (\ref{eq:3DHam}) should evolve as
\begin{equation} \begin{array}{c}
G_2(t) \propto \mbox{constant},\,\,\, G_3(t)
\propto \mbox{constant} ,\\ \\
G_4(t) \propto
\frac{1}{t^2},\,\,\,
G_5(t) \propto \frac{1}{t^4},\,\,\, G_6(t)
\propto \frac{1}{t^6}.
\end{array}\label{eq:3DHam_or_GALIs_approx}
\end{equation}
This behavior is seen for orbit R2 in Figs.~\ref{f:3d_reg_galis}(a)
and (c)
where the DOP853 and the TM method are used respectively for the
integration of the variational equations. Similarly to the case of
regular orbit R1 (Fig.~\ref{f:HH_reg_GALI}) the GALIs indicate that
the orbit is regular. On the other hand, in
Fig.~\ref{f:3d_reg_galis}(b) where the TDHeps method is applied, the
computed GALIs eventually show an exponential decay, wrongly
suggesting that orbit R2 is chaotic.

Finally, let us consider a particular regular orbit of the 8D
Hamiltonian system (\ref{eq:FPUHam}) which lies on a low dimensional
torus. In our study we impose fixed boundary conditions,
i.~e.~$q_0(t)=q_9(t)=p_0(t)=p_9(t)=0$ for all times $t$, fix the
system's parameter to $\beta=1.5$, and consider the regular orbit with
initial conditions $q_i=0.1$, $p_i=0$, $i=1,2,\ldots,8$, which we call
orbit R3. This orbit lies on a 4-dimensional torus and was also
studied in \cite{SBA08}.

\begin{figure}
\begin{tabular}{c}
  \includegraphics[width=\columnwidth]{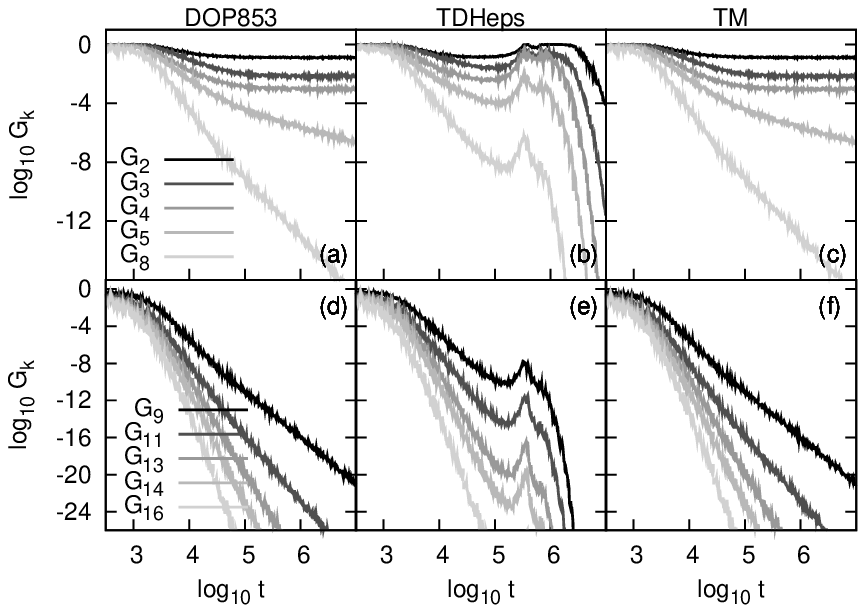}
\end{tabular}
\caption{The time evolution of $G_k(t)$, $k=2,3,4,5,8$ (upper
  panels) and $k=9,11,13,14,16$ (lower panels) for the regular orbit
  R3 of the 8D Hamiltonian system (\ref{eq:FPUHam}).  The variational
  equations are integrated by the DOP853 ((a) and (d)), the TDHeps
  ((b) and (e)), and the TM ((c) and (f)) method. The step size is
  $\tau=0.02$ for all methods. For the DOP853 method the parameter
  $\delta=10^{-5}$ is used.}
\label{f:fpu_galis}
\end{figure}

According to the theory of GALIs developed in \cite{SBA08}, regular
motion on a 4-dimensional torus implies that the corresponding
$G_2$, $G_3$ and $G_4$ remain practically constant, while the
remaining indices up to $G_{16}$ tend to zero following particular
power laws (see also Fig.~4 of \cite{SBA08}). As we can see from
Fig.~\ref{f:fpu_galis},
these expected behaviors are well reproduced when the DOP853
(Figs.~\ref{f:fpu_galis}(a) and (d)) and the TM
(Figs.~\ref{f:fpu_galis}(c) and (f)) methods are used for the
integration of the variational equations. On the other hand, the
TDHeps method fails to clearly determine the regular nature of orbit
R3, as well as the dimensionality of the torus on which the orbit
lies. From Figs.~\ref{f:fpu_galis}(b) and (e) we see that the computed
GALIs have a behavior similar to the one obtained by the DOP853 and
the TM methods, which indicates the regularity of the orbit, but only
up to $t\approx 10^5$. For $t> 10^5$ the computed GALIs eventually
show an exponential decay, wrongly suggesting that the orbit is
chaotic.

\section{Summary and discussion}
\label{sec:summary}

We considered the problem of the accurate and fast integration of the
variational equations of autonomous Hamiltonian systems. These
equations govern the evolution of a deviation vector from an orbit of
the system. The reliable determination of this evolution is necessary
when studies of the chaotic behavior of the system are needed. Many
chaos detection techniques, like the LCEs and the GALIs which we
considered in our study, are based on the evolution of one or more
deviation vectors.

We made a detailed presentation of several numerical schemes for the
integration of the variational equations and we applied them to
regular and chaotic orbits of Hamiltonian systems with different
number of degrees of freedom. We also investigated the efficiency of
these methods by comparing the CPU times they need for the computation
of the spectrum of LCEs, as well as their ability to accurately
reproduce well-known properties of the LCEs and the GALIs.

The evolution of deviation vectors cannot be separated from the
evolution of the orbit itself because the explicit expression of the
variational equations depend on the solution of the Hamilton's
equations of motion. Therefore, any general-purpose integration scheme
for ordinary differential equations, like the DOP853 integrator we
considered in our study, can be used for the simultaneous integration
of the set of equations which includes both the Hamilton's equations of
motion and the variational equations. This method proved to be very
reliable since it reproduced correctly the behavior of the LCEs and
the GALIs for all tested orbits and systems.

When the Hamiltonian function $H$ can be split into two integrable
parts $A$ and $B$, like $H=A+B$, symplectic integrators can be used
for the integration of the equations of motion. Symplectic integrators
are known to have better performance than non-symplectic ones for the
same integration time step, in terms of accuracy and required CPU
time. In order to investigate the applicability of such methods for
the integration of the variational equations, we focused our study
explicitly to Hamiltonians of the form $H=A+B$. In particular, we
considered Hamiltonians having a kinetic energy which is quadratic in
the momenta and a potential which depends only on the positions
(Eq.~(\ref{eq:HH_usual})). For such systems the two integrable parts
$A$ and $B$, are usually chosen to be the kinetic energy and the
potential respectively.  Most symplectic schemes require the
construction of symplectic maps $e^{\tau L_{A}}$ (\ref{eq:H_LA}) and
$e^{\tau L_{B}}$ (\ref{eq:H_LB}) for the solution of the integrable
parts $A$ and $B$. In our study we considered a very efficient
symplectic method, the $S_{B2}^c$ integrator, which has an extra degree
of complexity with respect to most symplectic integrators, since it
requires the explicit solution of an additional corrector term $C$
(map $e^{\tau L_{C}}$ (\ref{eq:H_LC})).

The variational equations of Hamiltonian (\ref{eq:HH_usual}) can be
written as the Hamilton's equations of motion of the time dependent
TDH (\ref{eq:HV_general}), whose coefficients are defined by the
coordinates of the orbit. Although individually the Hamilton's
equations of motion (\ref{eq:eq_mot_V}) and the variational equations
(\ref{eq:w_V}) are equations of motion of Hamiltonian functions, the
system (\ref{eq:L_HV}) which includes together both of them cannot be
considered as the equations of motion of a new generalized
Hamiltonian, and so, symplectic integrators cannot be directly used
for solving it. In our study we applied several approaches based
on symplectic techniques for the integration of the variational
equations. One approach we considered was the approximation of the
solution of the TDH through the knowledge of the orbit's coordinates
at specific times. These coordinates can be obtained by any symplectic
or non-symplectic integrator, independent of the method we use for
approximating the solution of the variational equations. In our study
we applied the $S_{B2}^c$ integrator for this purpose. First we assumed
the coefficients of the TDH to be constants for each integration step,
and we integrated the resulting quadratic TDH by the $S_{B2}^c$
integrator (TDHcc method) or solved it explicitely (TDHes method)
whenever this was possible (like for example in the case of the
H\'enon-Heiles system (\ref{eq:HH})). An alternative way we also
implemented was to use the $S_{B2}^c$ integrator for integrating the
time dependent TDH in an extended phase space (TDHeps method), using
again the knowledge of orbit's coordinates at specific times.  As an
application of the TDHeps method we refer to the numerical study of
the FPU problem in \cite{PP08} where a leap-frog integrator was used
for the integration of the time dependent TDH.

The TDHcc, TDHes and TDHeps methods had a rather poor numerical
performance as they failed in many cases to determine correctly the
regular or chaotic nature of orbits.  Our numerical results show that
the computed values of the LCEs cannot become smaller than a small
positive value, which sets a lower limit to the ability of these
techniques to numerically determine very small LCEs. So, one could
wrongly characterize regular orbits as slightly chaotic because their
computed LCEs cannot become smaller than the above-mentioned limit,
although their actual LCEs are zero. This happens for the regular
orbits R1 (Fig.~\ref{f:HH_reg_var}) and R2
(Fig.~\ref{f:3d_reg_lces}). Of course this limiting value decreases
for smaller integration steps because the numerical schemes
approximate better the real tangent dynamics of the system
(Fig.~\ref{f:dif_steps}(b)). Additionally, one could overestimate the
mLCE of chaotic orbits like for example in the case of the chaotic
orbit C2 (Fig.~\ref{f:C2_LCES}). Nevertheless these methods always
required less CPU time than the non-symplectic DOP853 method for the
same time step. Therefore these schemes can be used for some rough and
fast evaluation of LCEs' charts but not for the detailed investigation
of the dynamics or for the accurate computation of the LCEs and
GALIs. We note that among these three techniques the TDHeps method had
always the best numerical performance, although it required a bit more
CPU time than the other two methods.

The use of any symplectic scheme for the integration of the equations
of motion (\ref{eq:eq_mot_V}) of the $N$D Hamiltonian
(\ref{eq:HH_usual}) corresponds to the repeated action of a
$2N$-dimensional symplectic map $S$, constructed by the appropriate
composition of maps $e^{\tau L_{A}}$ (\ref{eq:H_LA}), $e^{\tau L_{B}}$
(\ref{eq:H_LB}) (and $e^{\tau L_{C}}$ (\ref{eq:H_LC}) if the corrector
term $C$ is used). Then, the tangent dynamics of the flow, i.~e.~the
solution of the variational equations (\ref{eq:w_V}), is described by
the tangent map $TS=\partial S/\partial \vec{x}$ of $S$ (some
particular implementations of this approach for different physical
problems can be found in \cite{TM_example1,TM_example2}).

The TM method we presented in our study provides a simple, systematic
technique to construct the tangent map $TS$ for any general symplectic
integration scheme used for the integration of the orbit, which is
perfectly suited for practical implementations. According to this
method, one has to substitute the $2N$-dimensional maps $e^{\tau
  L_{A}}$ (\ref{eq:H_LA}), $e^{\tau L_{B}}$ (\ref{eq:H_LB}), $e^{\tau
  L_{C}}$ (\ref{eq:H_LC}) needed for the symplectic integration of the
equations of motion (\ref{eq:eq_mot_V}), by the extended
$4N$-dimensional maps $e^{\tau L_{AV}}$ (\ref{eq:H_LAV}), $e^{\tau
  L_{BV}}$ (\ref{eq:H_LBV}), $e^{\tau L_{CV}}$ (\ref{eq:H_LCV})
respectively. This procedure leads to the construction of an extended
$4N$-dimensional final map composed by the $2N$-dimensional maps $S$
and $TS$.  In particular, the first $2N$ equations of this map are the
equations of map $S$, and the rest $2N$ equations form the tangent map
$TS$.

The TM method and the DOP853 integrator were the only techniques that
succeeded in computing correctly the LCEs and the GALIs for all
studied cases. Among them, the TM method required  less
CPU time for the same integration step size. Another advantage of the
TM method over the DOP853 integrator is that its application with
larger time steps reduces the needed CPU time, keeps the accuracy to
acceptable levels, and produce more reliable results than the DOP853
integrator.

In conclusion, the TM method proved to be the most efficient one among
all tested methods, since it required the least CPU time for the
computation of the spectrum of LCEs and reproduced very accurately the
behavior of the LCEs and GALIs. Therefore, whenever the studied
Hamiltonian can be split into two integrable parts, so that it can be
integrated by symplectic integrators, the TM method should be
preferred over other symplectic or non-symplectic integration schemes.

Although we considered in our study applications of the TM method to
Hamiltonian systems of relatively low dimensionality (systems having
up to eight degrees of freedom), the method is expected to be also
very efficient for higher-dimensional systems. Symplectic integrators
have already been applied successfully for the accurate integration of
motion in multi-dimensional systems which are related for example, to
problems of astronomical interest (e.~g.~\cite{TM_example2}), of
molecular dynamics (e.~g.~\cite{GNS94,OMF_23}) and dynamics of
nonlinear lattices (e.~g.~\cite{si_appl2}).  Using the TM method these
symplectic integration schemes can be extended to integrate also the
corresponding variational equations. This is a problem of great
practical importance, which we plan to address in a future
publication.

As a final remark, we note that all the presented methods require the
knowledge of the analytic expression of matrix
$\textbf{D$^2_V$}(\vec{q}(t))$ (\ref{eq:D2V}) (or of matrix
$\textbf{A}(t)$ (\ref{eq:var}) in the case of a general dynamical
system). If the variational equations cannot be written explicitly,
possibly due to the complicated form of the studied dynamical system,
the analytical derivation of these matrices is not possible and their
elements could be estimated numerically, introducing an additional
error to the solution of the variational equations. An approach that
could be followed in such cases is the approximation of the solution
of the variational equations by the difference of two orbits initially
located very close to each other (see \cite{two_particles} for some
particular applications of this approach). This is the so-called
\textit{two-particle method}, which was introduced in \cite{BGS76} and
is mainly used for the evaluation of the mLCE. It was realized almost
immediately after the introduction of this technique that this
approach is less efficient and reliable than the actual integration of
the variational equations  \cite{CGG78} (whenever, of course, this integration is
possible). For this reason we did not include this
approach in our study.

\begin{acknowledgments}
  Ch.~S. would like to thank G.~Benettin, J. Bodyfelt, T.~Bountis,
  T.~Kov\'{a}cs, J.~Laskar and A.~Ponno for useful
  discussions. E.~G. acknowledges financial support by the DFG project
  SO 216/21-1. We would also like to thank the anonymous referees for
  very useful comments and suggestions which helped us improve the
  clarity of the paper.
\end{acknowledgments}

\appendix
\section{Analytical expressions for the integration of the
  H\'enon-Heiles system}

We present here the explicit expressions of the various
integration schemes for the 2D H\'enon-Heiles system, whose Hamiltonian
function (\ref{eq:HH}) is of the form (\ref{eq:HH_usual}) with
$\vec{q}=(x,y)$, $\vec{p}=(p_x,p_y)$. The Hamilton's equations of motion
(\ref{eq:eq_mot_V}) are
\begin{equation}
\begin{array}{ccl}
\dot{x}&=& p_x \\
\dot{y}&=& p_y \\
\dot{p}_x&=& -x-2xy \\
\dot{p}_y&=& y^2-x^2-y
\end{array}.
\label{eq:HHeq}
\end{equation}
The variational equations (\ref{eq:w_V}) of the system are
\begin{equation}
\begin{array}{ccl}
\dot{\delta x}&=& \delta p_x \\
\dot{\delta y}&=& \delta p_y \\
\dot{\delta p}_x&=& -(1+2y) \delta x -2x \delta y \\
\dot{\delta p}_y&=&  -2x \delta x + (-1+2y) \delta y
\end{array},
\label{eq:HH_var_eq}
\end{equation}
while the corresponding TDH (\ref{eq:HV_general}) takes the form
\begin{equation}
\begin{array}{c}
  \displaystyle
  H_{VH}(\delta x,\delta y,\delta p_x, \delta p_y;t)= \frac{1}{2} \left(\delta
    p_x^2+\delta p_y^2\right)+ \\ \\   + \displaystyle \frac{1}{2} \left\lbrace
    \left[1+2y(t)\right]\delta x^2 +  \left[1-2y(t)\right] \delta y^2 +
    2\left[2x(t) \right] \delta x \delta y\right\rbrace.
\end{array}
\label{eq:HH_var}
\end{equation}

\subsection{Symplectic integration of the  equations of motion}
\label{ap:HH}

The H\'enon-Heiles Hamiltonian (\ref{eq:HH}) can be split into two
parts $H_2 = A+B$, according to equation (\ref{eq:H_AB}), with
\begin{equation}
\begin{array}{lll}
A &  = & \displaystyle  \frac{1}{2} (p_x^2+p_y^2),\\ \\ B & = &
 \displaystyle \frac{1}{2} (x^2+y^2) + x^2 y -
\frac{1}{3} y^3.
\end{array}
\label{eq:ham_equiv}
\end{equation}
As it was explained in section \ref{sec:SI}, this separation is
convenient for the application of symplectic schemes for the
integration of equations (\ref{eq:HHeq}), since Hamiltonians $A$ and
$B$ are integrable.  The maps $e^{\tau L_A}$ (\ref{eq:H_LA}), $e^{\tau
  L_B}$ (\ref{eq:H_LB}), which propagate the set of initial conditions
$(x,y,p_x,p_y)$ at time $t$, to their final values $(x',y',p_x',p_y')$
at time $t+\tau$ are
\begin{equation}
e^{\tau L_A}: \left\{ \begin{array}{lll}
x' & = & x + p_x \tau \\
y' & = & y + p_y \tau \\
p_x' & = & p_x \\
p_y' & = & p_y
\end{array}\right. ,
\label{eq:HH_LA}
\end{equation}
\begin{equation}
e^{\tau L_B}: \left\{ \begin{array}{lll}
x' & = &  x\\
y' & = &  y\\
p_x' & = & p_x - x(1+2y) \tau \\
p_y' & = & p_y + (y^2-x^2 -y) \tau
\end{array}\right. .
\label{eq:HH_LB}
\end{equation}
The corrector
term (\ref{eq:H_ABB}) is
\begin{equation}
C=\{B,\{B,A\}\}=\left( x+2xy \right)^2 + \left( x^2 -y^2 +y \right)^2   ,
\label{eq:HH_ABB}
\end{equation}
and the corresponding map  $e^{\tau L_C}$ (\ref{eq:H_LC})
\begin{equation}
e^{\tau L_C}: \left\{ \begin{array}{lll}
x' & = &  x\\
y' & = &  y\\
p_x' & = & p_x - 2x(1+2x^2+6y+2y^2) \tau \\
p_y' & = & p_y - 2(y-3y^2+2y^3+3x^2+2x^2y) \tau
\end{array}\right. .
\label{eq:HH_LC}
\end{equation}

\subsection{Integration of the variational equations}
\label{ap:HH_var_eq_all}

We derive now for the particular case of the H\'enon-Heiles system the
analytical expressions of the various numerical schemes presented in
section \ref{sec:num_int} for the
integration of the variational equations.

\subsubsection{Diagonal form of the TDH (\ref{eq:HH_var}) with
constant coefficients} \label{ap:transform}

Inserting the values $x(t_i)\equiv x_i$, $y(t_i)\equiv y_i$ at a
specific time $t_i$ in the functional form of the TDH
(\ref{eq:HH_var}), $H_{VH}$ becomes a quadratic 2D Hamiltonian with
constant coefficients. The equations of motion of this Hamiltonian are
solved immediately if $x_i=0$. For $x_i \neq 0$ the transformation
\begin{equation}
\left[ \begin{array}{c}
\delta x \\ \delta y
       \end{array} \right] = \mathbf{T} \left[ \begin{array}{c}
\delta X \\ \delta Y
       \end{array} \right] \,\,\, , \,\,\,
\left[ \begin{array}{c}
\delta p_x \\ \delta p_y
       \end{array} \right] = \mathbf{T} \left[ \begin{array}{c}
\delta P_X \\ \delta P_Y
       \end{array} \right]
\label{eq:trans_HH}
\end{equation}
with
\begin{widetext}
\begin{equation}
\mathbf{T} = \left[ \begin{array}{cc} \displaystyle \frac{\sqrt{x^2_i+y^2_i +
y_i\sqrt{x^2_i+y^2_i} }} {\sqrt{2} \sqrt{x^2_i+y^2_i}} & \displaystyle
\frac{-x_i}{\sqrt{2} \sqrt{x^2_i+y^2_i + y_i\sqrt{x^2_i+y^2_i} }} \\
\displaystyle \frac{x_i \sqrt{x^2_i+y^2_i + y_i\sqrt{x^2_i+y^2_i} }}{\sqrt{2}
\sqrt{x^2_i+y^2_i} \left(\sqrt{x^2_i+y^2_i} +y_i \right) } & \displaystyle
\frac{\sqrt{x^2_i+y^2_i} +y_i}{\sqrt{2} \sqrt{x^2_i+y^2_i +
y_i\sqrt{x^2_i+y^2_i} }}
\end{array} \right],
\label{eq:T_HH}
\end{equation}
\end{widetext}
gives $H_{VH}(\delta x,\delta y,\delta p_x, \delta p_y;t_i)$ the
diagonal form
\begin{equation}
\begin{array}{c}
\displaystyle H_{VHD}(\delta X,\delta Y,\delta P_x, \delta P_y)= \frac{1}{2}
\left(\delta P_x^2+\delta P_y^2\right)+ \\ \\ + \displaystyle \frac{1}{2}
\left\lbrace \left( 1+2\sqrt{x^2_i+y^2_i} \right) \delta X^2 + \left(
1-2\sqrt{x^2_i+y^2_i}\right) \delta Y^2 \right\rbrace.
\end{array}
\label{eq:HH_var_diag}
\end{equation}
The columns of matrix $\mathbf{T}$ are the
eigenvectors of matrix
\begin{equation}
\displaystyle \mathbf{D}^2_V(\vec{q}(t_i)) \equiv
\mathbf{D}^2_B (x_i,y_i)=\left[
\begin{array}{cc} \displaystyle
1+2y_i & 2x_i
\\
2x_i & 1-2y_i
\end{array}
\right] ,
\label{eq:HH_B_mat}
\end{equation}
and $\lambda_{1,2}=1\pm 2 \sqrt{x^2_i+y^2_i}$ the corresponding
eigenvalues.

\subsubsection{Symplectic integration of the TDH (\ref{eq:HH_var})
in an extended phase space}
\label{EPS_HH}

Considering the TDH (\ref{eq:HH_var}) as a time dependent Hamiltonian,
we can transform it to a time independent one having time $t$ as an
additional generalized position by the procedure presented in section
\ref{sec:EPS}.  The 3D Hamiltonian (\ref{eq:HV_tilde_general}) takes
the form
\begin{equation}
\begin{array}{c}
\displaystyle \widetilde{H}_{VH}(\delta x,\delta y,t,\delta p_x, \delta p_y,
p_t)= \frac{1}{2} \left(\delta p_x^2+\delta p_y^2\right)+ p_t \\ \\ +
\displaystyle \frac{1}{2} \left\lbrace \left[1+2y(t)\right]\delta x^2 +
\left[1-2y(t)\right] \delta y^2 + 2\left[2x(t) \right] \delta x \delta
y\right\rbrace,
\end{array}
\label{eq:HH_var_EPS}
\end{equation}
with $p_t$ being the conjugate momentum of coordinate
$t$. $\widetilde{H}_{VH}$ can be split into two integrable parts
(\ref{eq:HV_tilde_AB})
\begin{equation}
\begin{array}{lll}
\widetilde{A}(\delta p_x, \delta p_y, p_t)& = & \displaystyle \frac{1}{2}
\left(\delta p_x^2+\delta p_y^2\right)+ p_t,\\ \\ \widetilde{B}( \delta
x,\delta y,t) & = & \displaystyle \frac{1}{2} \left\lbrace
\left[1+2y(t)\right]\delta x^2 + \left[1-2y(t)\right] \delta y^2 + \right. \\
\\ & & + \left. \displaystyle 2\left[2x(t) \right] \delta x \delta
y\right\rbrace,
\end{array}
\label{eq:HVH_tilde_AB}
\end{equation}
so that its equations of motion can be integrated by any symplectic
integration method in order to obtain the time evolution of variations
$\delta x$, $\delta y$, $\delta p_x$, $\delta p_y$. The maps $e^{\tau
  L_{\widetilde{A}}}$ (\ref{eq:HV_tilde_LA}), $e^{\tau
  L_{\widetilde{B}}}$ (\ref{eq:HV_tilde_LB}) (neglecting the equations
for $p_t$) are
\begin{equation}
e^{\tau L_{\widetilde{A}}}: \left\{ \begin{array}{lll}
\delta x' & = &  \delta x + \delta p_x \tau\\
\delta y' & = &  \delta y + \delta p_y \tau\\
t' & = & t + \tau\\
\delta p_x' & = & \delta p_x \\
\delta p_y' & = & \delta p_y
\end{array}\right. ,
\label{eq:HVH_tilde_LA}
\end{equation}
\begin{equation}
e^{\tau L_{\widetilde{B}}}: \left\{ \begin{array}{lll} \delta x' & = & \delta
x \\ \delta y' & = & \delta y \\ t' & = & t\\ \delta p_x' & = & \delta p_x -
\left\lbrace \left[ 1+2y(t)\right] \delta x +2x(t) \delta y \right\rbrace \tau
\\ \delta p_y' & = & \delta p_y + \left\lbrace -2x(t) \delta x + \left[
-1+2y(t)\right] \delta y \right] \tau
\end{array}\right. .
\label{eq:HVH_tilde_LB}
\end{equation}
The corrector term $\widetilde{C}$ (\ref{eq:HV_tilde_ABB}) is
\begin{equation}
\widetilde{C}=\left[ \delta x + 2 x(t) \delta y +2 y(t) \delta x \right]^2 +
\left[ \delta y + 2 x(t) \delta x -2 y(t) \delta y \right]^2 ,
\label{eq:HVH_tilde_ABB}
\end{equation}
and the corresponding map $e^{\tau L_{\widetilde{C}}}$ (\ref{eq:HV_tilde_LC})
\begin{equation}
e^{\tau L_{\widetilde{C}}}: \left\{ \begin{array}{lll} \delta x' & = & \delta
x \\ \delta y' & = & \delta y \\ t' & = & t\\ \delta p_x' & = & \delta p_x - 2
\left\lbrace 4 x(t) \delta y + \right. \\ & & \left. + \left[ 4 x^2(t) +
\left( 1+2y(t) \right)^2 \right]\delta x \right\rbrace \tau \\ \delta p_y' & =
& \delta p_y - 2 \left\lbrace 4 x(t) \delta x + \right. \\ & & \left. + \left[
4 x^2(t) + \left( 1-2y(t) \right)^2 \right]\delta y \right\rbrace \tau
\end{array}\right. .
\label{eq:HVH_tilde_LC}
\end{equation}

\subsubsection{The tangent map method}
\label{TMM_HH}

According to the TM method presented in section \ref{sec:TMM_gen}
equations (\ref{eq:HHeq}) and (\ref{eq:HH_var_eq}) form a set of
equations which defines the act of the differential operator $L_{HV}$
on vector $\vec{u}=(x,y,p_x,p_y,\delta x,\delta y,\delta p_x,\delta
p_y)$ (equations (\ref{eq:L_HV})). This set of equations is split into
two integrable sets
\begin{equation}
\left.
\begin{array}{ccl}
\dot{x}&=& p_x \\
\dot{y}&=& p_y \\
\dot{p}_x&=& 0 \\
\dot{p}_y&=& 0 \\
\dot{\delta x}&=& \delta p_x \\
\dot{\delta y}&=& \delta p_y \\
\dot{\delta p}_x&= &0 \\
\dot{\delta p}_y&=&  0
\end{array} \right\} \Rightarrow \frac{d \vec{u}} {dt}= L_{AV}
\vec{u},
\label{eq:HH_L_AV}
\end{equation}
\begin{equation}
\left.
\begin{array}{ccl}
\dot{x}&=& 0 \\
\dot{y}&=& 0 \\
\dot{p}_x&=& -x-2xy \\
\dot{p}_y&=& y^2-x^2-y \\
\dot{\delta x}&=& 0 \\
\dot{\delta y}&=& 0 \\
\dot{\delta p}_x&=& -(1+2y) \delta x -2x \delta y \\
\dot{\delta p}_y&=&  -2x \delta x + (-1+2y) \delta y
\end{array} \right\} \Rightarrow \frac{d \vec{u}} {dt}= L_{BV}
\vec{u}, \label{eq:HH_L_BV}
\end{equation}
which define the act of operators $L_{AV}$ (\ref{eq:L_AV}) and
$L_{BV}$ (\ref{eq:L_BV}) respectively. Then, maps $e^{\tau L_{AV}}$
(\ref{eq:H_LAV}) and $e^{\tau L_{AV}}$ (\ref{eq:H_LBV}) are
\begin{equation}
e^{\tau L_{AV}}: \left\{ \begin{array}{lll}
x' & = &  x + p_x \tau\\
y' & = &  y + p_y \tau\\
px' & = & p_x \\
py' & = & p_y \\
\delta x' & = &  \delta x + \delta p_x \tau\\
\delta y' & = &  \delta y + \delta p_y \tau\\
\delta p_x' & = & \delta p_x \\
\delta p_y' & = & \delta p_y
\end{array}\right. ,
\label{eq:HH_H_LAV}
\end{equation}
\begin{equation}
e^{\tau L_{BV}}: \left\{ \begin{array}{lll}
x' & = &  x\\
y' & = &  y\\
p_x' & = & p_x - x(1+2y) \tau \\
p_y' & = & p_y + (y^2-x^2 -y) \tau \\
\delta x' & = & \delta x \\
\delta y' & = & \delta y \\
\delta p_x' & = & \delta p_x - \left[ (1+2y) \delta x +2x \delta y \right] \tau \\
\delta p_y' & = & \delta p_y + \left[ -2x \delta x + (-1+2y)
\delta y \right] \tau
\end{array}\right. ,
\label{eq:HH_H_LBV}
\end{equation}
while the map $e^{\tau L_{CV}}$ (\ref{eq:H_LCV})  of the corrector
function $C$ (\ref{eq:HH_ABB}) is
\begin{equation}
e^{\tau L_{CV}}: \left\{ \begin{array}{lll}
x' & = &  x\\
y' & = &  y\\
p_x' & = & p_x - 2x(1+2x^2+6y+2y^2) \tau \\
p_y' & = & p_y - 2(y-3y^2+2y^3+3x^2+2x^2y) \tau \\
\delta x' & = & \delta x \\
\delta y' & = & \delta y \\
\delta p_x' & = & \delta p_x - 2 \left[ (1+6x^2+2y^2+6y) \delta x + \right. \\
& & \left. + 2x(3+2y) \delta y \right] \tau \\
\delta p_y' & = & \delta p_y -2 \left[ 2x(3+2y) \delta x + \right. \\
& & \left. + (1+2x^2+6y^2-6y) \delta y \right] \tau
\end{array}\right.  .
\label{eq:HH_H_LCV}
\end{equation}



\begin{thebibliography}{99}
\bibitem{comment1} Note that in \cite{LR01} the Poisson bracket
  (\ref{eq:Poisson}) is defined with opposite signs.
\bibitem{comment2} In \cite{LR01} this commutator is denoted as
  $C=\{\{A,B\},B\}$ due to the different used definition of the
  Poisson bracket (\ref{eq:Poisson}).
\end{thebibliography}
\end{document}